\documentclass[journal,draftclsnofoot,onecolumn,12pt]{IEEEtran}
\ifCLASSINFOpdf
\else
\fi

\usepackage{graphicx}%插入图片
\usepackage{bm}%粗体字包
\usepackage{amsfonts}
\usepackage{amsmath}
\usepackage{amssymb}
\usepackage{times}
\usepackage{subfigure}
\usepackage{latexsym,bm,amsmath,amssymb} %Latex 中数学包
\usepackage{CJK}
\usepackage{hhline}

\usepackage{multirow} %multirow for format of table
\usepackage{amsmath}
\usepackage{xcolor}
\usepackage[linesnumbered,ruled,vlined]{algorithm2e}
\usepackage{epstopdf}
\usepackage{cite}
\usepackage{amssymb}
\usepackage{bbding}
\usepackage{pdfrender}
\newcommand*{\boldcheckmark}{%
  \textpdfrender{
    TextRenderingMode=FillStroke,
    LineWidth=.5pt, % half of the line width is outside the normal glyph
  }{\checkmark}%
}
\begin{document}

\title{Faster-than-Nyquist Asynchronous NOMA Outperforms Synchronous NOMA}

\author{Shuangyang~Li,~\IEEEmembership{Student Member,~IEEE,}  Zhiqiang~Wei,~\IEEEmembership{Member,~IEEE,} Weijie~Yuan,~\IEEEmembership{Member,~IEEE,} Jinhong~Yuan,~\IEEEmembership{Fellow,~IEEE,}
Baoming~Bai,~\IEEEmembership{Senior Member,~IEEE}, Derrick Wing Kwan Ng,~\IEEEmembership{Fellow,~IEEE}, and Lajos~Hanzo,~\IEEEmembership{Fellow,~IEEE}
\vspace{-15mm}
\thanks{
   Part of the paper was presented in the IEEE Wireless Communications and Networking Conference 2021~\cite{ShuangyangWCNC}.
  }
}

\maketitle

\begin{abstract}
Faster-than-Nyquist (FTN) signaling aided non-orthogonal multiple access (NOMA) is conceived and its achievable rate is quantified in the presence of \emph{random} link delays of the different users. We reveal that exploiting the link delays may potentially lead to a signal-to-interference-plus-noise ratio (SINR) gain, while transmitting the data symbols at FTN rates has the potential of increasing the degree-of-freedom (DoF).
We then unveil the fundamental trade-off between the SINR and DoF. In particular, at a sufficiently high symbol rate, the SINR gain vanishes while the DoF gain achieves its maximum, where the achievable rate is almost $(1+\beta)$ times higher than that of the conventional synchronous NOMA transmission in the high signal-to-noise ratio (SNR) regime, with $\beta$ being the roll-off factor of the signaling pulse.
Our simulation results verify our analysis and demonstrate considerable rate improvements over the conventional power-domain NOMA scheme.
\end{abstract}


\IEEEpeerreviewmaketitle

%\linespread{1.1}
\section{Introduction}
The escalating number of wireless devices and sensors has inspired exploring efficient multiple access solutions for future Internet-of-things (IoT) networks.
In particular, non-orthogonal multiple access (NOMA) has been extensively studied as a promising radio access scheme due to its superiority in handling massive connectivity \cite{ding2017survey}, which is a crucial requirement for IoT networks.
In contrast to conventional orthogonal multiple access (OMA) schemes, NOMA allows more than one user to transmit their information via the same time-frequency resource block, hence enjoying an improved spectral efficiency via exploiting their channel disparities~\cite{WeiPerformanceGain}.
However, future beyond fifth-generation (B5G) wireless networks are expected to attain even higher spectral efficiency than the 5G networks~\cite{Wu2020Towards}. Therefore, the existing NOMA schemes have to be improved in order to meet the ultra-high data rate requirement of B5G wireless networks.

Hence, diverse advanced transmission schemes have been proposed in the literature~\cite{wang2021minimum,rimoldi1996rate,Clerckx2016rate,Ding2020simple_IRS_NOMA,Moltafet2018comparison,HaciAsyn,Zou2019analysis,Yuan2020iterative}.
For example, a NOMA-based system design by considering minimum error probability was conceived in~\cite{wang2021minimum}, where the
effect of realistic imperfect successive interference cancellation (SIC) was also taken into account for the system design.
The simulation results showed that a superior error performance can be obtained based on the proposed design compared to the conventional power-domain NOMA.
Furthermore, the rate-splitting multiple access (RSMA) was considered in~\cite{rimoldi1996rate}, where each user partially decoded the interference and treated the rest of the interference as noise. In this way, RSMA has been shown, by Clerckx \emph{et. al.}, to achieve both an improved spectral efficiency and an enhanced robustness against imperfect channel state information at the transmitter~\cite{Clerckx2016rate}.
Besides, an intelligent reflecting surface (IRS) assisted NOMA scheme was designed in~\cite{Ding2020simple_IRS_NOMA} by Ding \emph{et. al.}, where the IRSs are used for beneficially aligning the cell-edge users' effective channel vectors with the predetermined spatial directions in order to improve the overall performance of NOMA transmissions.
Moreover, the sparse code multiple access (SCMA) transmission allows more users to transmit their information among a less number of resource blocks using an appropriately designed codebook~\cite{Moltafet2018comparison}. By relying on a carefully constructed codebook, it has been shown that SCMA can achieves a better error performance than conventional power-domain NOMA~\cite{Moltafet2018comparison}.
On the other hand, asynchronous NOMA (aNOMA) transmissions have also been shown to offer an extended achievable rate region compared to conventional synchronous NOMA transmissions~\cite{HaciAsyn,Zou2019analysis,ShuangyangWCNC}.
Specifically, Haci \emph{et. al.}~\cite{HaciAsyn} conceived the aNOMA concept based on orthogonal frequency-division multiplexing (OFDM) and showed that the aNOMA scheme outperforms both the conventional synchronous NOMA scheme and the classic orthogonal frequency-division multiple access (OFDMA) arrangement in terms of both its bit error rate (BER) and capacity.
Furthermore, time domain (TD) aNOMA transmission was evaluated by Zou \emph{et. al.}~\cite{Zou2019analysis}, where the authors considered a fixed link delay introduced by each user and the data was conveyed by finite-duration signaling pulses. Similar to the frequency domain (FD) aNOMA transmission, the TD aNOMA technique has also shown advantages in terms of its achievable rate~\cite{Zou2019analysis}.
Moreover, NOMA transmission based on faster-than-Nyquist (FTN) signaling was also considered in the literature~\cite{Yuan2020iterative}. Indeed, FTN signaling is a classic non-orthogonal signaling scheme exhibiting an enhanced spectral efficiency~\cite{General-anderson2013faster,li2020code,sugiura2014frequency,ishihara2021evolution}, where data is transmitted at a symbol rate higher than the Nyquist intersymbol interference (ISI)-free rate{\footnote{It should be noted that the idea of transmitting symbols faster is not limited to the TD. For example, it has been shown that transmitting symbols faster in the FD also enjoys advantages~\cite{kanaras2009spectrally,darwazeh2013optical}.}}~\cite{li2018superposition}. Although FTN-based NOMA (FTN-NOMA) transmission is expected to have an increased achievable rate, its theoretical rate analysis has not been disseminated in the open literature.

As an extension of the state-of-the-art, we improve the conventional power-domain NOMA scheme for bandlimited uplink transmissions from two different perspectives, namely, its signal-to-interference-plus-noise ratio (SINR) and its degree-of-freedom (DoF). In particular, we show that an SINR gain can be achieved by exploiting the \emph{link delay difference} between different users, while a DoF gain is attained by FTN signaling. However, interestingly enough, there is a fundamental trade-off between the SINR gain and DoF gain.
In fact, the potential of both the SINR and DoF improvements arise from the spectral aliasing of signal transmissions under a practical bandwidth constraint. Apart from the practical benefits of an enhanced throughput, the proposed NOMA scheme also provides theoretical insights. It is widely recognized  that spectral aliasing
occurs when the signal bandwidth is higher than half the sampling frequency. We reveal two ways of mitigating the effect of spectral aliasing. The first one is by imposing phase rotations on the signal spectrum (corresponding to imposing delay for the TD signal), so that the power spectral density (PSD) at the aliased FD components generally has a reduced value due to the superposition of adjacent spectra (caused by spectral aliasing).
The second one is to transmit signals at an FTN symbol rate\footnote{It should be noted that the potential DoF gain is not due to the oversampling but owing to the increased symbol rate at the transmitter side. More details on this argument can be found in~\cite{rusek2009constrained}.}. These two solutions lead to potential SINR and DoF gains, and we will show that there is a trade-off between these gains.

To further clarify the above arguments, we conceive more complex but practical uplink transmission scenarios in this paper, where each user experiences a \emph{random} link delay.
This is in contrast to the majority of the literature, e.g., \cite{ding2017survey,Dai2018survey} and the references therein, which assumed perfectly time-synchronous transmission among NOMA users. However, the perfect synchronization assumption is often unrealistic in practical uplink transmissions, due to the use of different clock generators adopted at geographically distributed uplink users and owing to the distance-dependent propagation delays. As a consequence, the signals from multiple users cannot be synchronously superimposed at the BS as commonly assumed in the literature{\footnote{Although the conception of the 2G system's adaptive time-frame alignment has been used, in the face of mobility, non-negligible time-of-arrival differences are experienced at the uplink receiver.}} \cite{WeiPerformanceGain,ding2017survey,Dai2018survey}.
Given practical considerations, we study the achievable rates of both the asynchronous FTN-NOMA (aFTN-NOMA) scheme and the aNOMA scheme (aFTN-NOMA scheme using Nyquist symbol rate). We also compare them to the achievable rates of the conventional synchronous power-domain NOMA
schemes under the same channel conditions. %In particular, we derive the closed-form expression of the achievable rates under the successive interference cancellation (SIC) detection and invoke Szeg\"o's Theorem~\cite{gray2006toeplitz} to obtain some important insights on different NOMA schemes.
In summary, our work is motivated by the following facts: 1) Although FTN-NOMA transmission has been proposed in the literature, its transmission with arbitrary link delay has not been considered, even though it is of great practical interest; 2) The theoretical rate analysis of FTN-NOMA is absent in the literature and the
potential rate improvement attained by asynchronous transmissions due to arbitrary link delays is still not well-understood; 3) The fundamental relationships between the symbol rate, asynchronous transmission, and spectrum aliasing has not been unveiled previously. Furthermore, their effects on the achievable rate has not been documented. For a better understanding of our motivations and the novelty of this work, we have summarized the major contributions of the related literature in Table~\ref{Novelity_contributions} in comparison to this work.
\begin{table}[htbp]
\caption{Summary of Related Works}
\centering
\begin{tabular}{|l|c|c|c|c|c|}
\hline
Related works~&~\cite{ShuangyangWCNC}~&~\cite{HaciAsyn}~&~\cite{Zou2019analysis}~&~\cite{Yuan2020iterative}~&~This work \\
\hline
Asynchronous transmission~&~\boldcheckmark ~&~\boldcheckmark ~&~\boldcheckmark ~&~\XSolidBrush~&~\boldcheckmark \\
\hline
Arbitrary link delay~&~\boldcheckmark ~&~\XSolidBrush~&~\XSolidBrush~&~\XSolidBrush~&~\boldcheckmark  \\
\hline
FTN signaling~&~\XSolidBrush~&~\XSolidBrush~&~\XSolidBrush~&~\boldcheckmark ~&~\boldcheckmark \\
\hline
Achievable rate analysis~&~\boldcheckmark ~&~\boldcheckmark ~&~\boldcheckmark  ~&~\XSolidBrush~&~\boldcheckmark  \\
\hline
\end{tabular}
\label{Novelity_contributions}
\end{table}
Corresponding to the motivations of this work, the main contributions of this paper are summarized as follows:
\begin{itemize}
\item We derive the closed-form expression of the mutual information for aFTN-NOMA schemes at various symbol rates using SIC detection and specific link delays for each user.
\item In order to characterize the effect of link delays, we propose to apply bounding techniques to the mutual information derived for each user. Specifically, we show that the link delay may potentially affect the achievable rate via changing the discrete-time Fourier transform (DTFT) of the transmitted signal. Therefore, we derive both the upper- and lower-bounds of the signal DTFTs that are independent from the link delays.
\item Based on the DTFT bounds derived, we invoke Szeg\"o's Theorem~\cite{gray2006toeplitz} for characterizing the mutual information of the aFTN-NOMA schemes at various symbol rates and derive the corresponding upper- and lower-bounds of the achievable rates. Based on the bounds derived, the influence of the link delay and symbol rate on the signal spectrum (DTFT) is unveiled. In particular,
    we also reveal the relationships between the effect of spectral aliasing and the link delay as well as the symbol rate.
\item We show that exploiting the link delay actually results in an SINR gain, while increasing the symbol rate may lead to a DoF gain, but there is a trade-off between them. Explicitly, at a sufficiently high symbol rate, the SINR gain vanishes, while the DoF gain attains its maximum. In this case, the achievable rate is essentially the capacity associated with the specific signaling pulse. Explicitly, this is about $1+\beta$ times higher than the conventional synchronous NOMA scheme in the high signal-to-noise ratio (SNR) regime, with $0 \le \beta \le 1$ being the FD roll-off factor of the signaling pulse.
\item Simulation results confirm the accuracy of our analysis, and demonstrate a significant improvement in terms of the rate attained by the aFTN-NOMA scheme compared to the perfectly synchronous NOMA scheme.

\end{itemize}

\emph{Notations:} $\max \left\{ {\cdot} \right\}$ and $\min \left\{ {\cdot} \right\}$ denote the maximization and minimization operations, respectively; $ \otimes $ denotes the convolution operation; $I\left( \cdot ; \cdot \right)$ and $h\left( \cdot \right)$ denote the mutual information and the differential entropy, respectively; %$I\left( \cdot ; \cdot |\cdot \right)$ and $h\left( \cdot |\cdot \right)$ denote the conditional mutual information and the conditional differential entropy, respectively;
$\delta(\cdot)$ denotes the Dirac delta function; ${{\bf I}_{N \times N}}$ denotes the identity matrix of size $N \times N$; the notations $(\cdot)^{\rm{T}}$, $(\cdot)^{*}$ represent the transpose and the conjugate operations for a matrix, respectively; the blackboard bold letter ${\mathbb E}\left[ {\cdot} \right]$, and ${\mathbb C}$
denote the expectation operator and the complex number field, respectively.

\section{System Model}
Let us consider a pair of single-carrier asynchronous NOMA uplink schemes transmitting over block-fading channels. Specifically, we assume that there are $K$ users and each user transmits $N$ information symbols, i.e., ${{\bf{x}}_k} = {\left[ {{x_k}\left[ 0 \right],{x_k}\left[ 1 \right], \ldots, {x_k}\left[ {N - 1} \right]} \right]^{\rm{T}}}$, for $ 1 \le k \le K$.
To model the asynchronous transmissions, we assume a \emph{random} link delay $\tau \left[ k \right], 1 \le k \le K$, for each user.
We adopt the common block-fading channel model of NOMA systems. Let $d_k$ denote the distance between the $k$-th user and the base station (BS) and $\alpha$ represent the path loss factor.
Then, the channel coefficient ${h_k\in {\mathbb C}}, 1 \le k \le K$, of the $k$-th user is assumed to be complex-valued Gaussian distributed with a zero mean and
variance of $\frac{1}{{1 + d_k^\alpha }}$~\cite{WeiPerformanceGain,ding2014performance}.
Let $T$ denote the Nyquist symbol duration.
Without loss of generality, we consider the root raised cosine (RRC) pulse having an FD roll-off factor $0 \le \beta \le 1$ as our signaling pulse $p(t)$, which is bandlimited, real-valued, and $T$-orthogonal with a normalized energy, i.e., $\int_{ - \infty }^\infty  {{{\left| {p\left( t \right)} \right|}^{\rm{2}}}} {\rm{d}}t = 1$.
The Fourier transform of $p(t)$ is denoted by ${{H_p}\left( f \right)}$, which is strictly bandlimited within the frequency interval of $f \in \left[ { - W,W} \right]$, with $W \buildrel \Delta \over = \frac{{1 + \beta }}{{2T}}$ denoting the baseband bandwidth.
In what follows, we will present the system models of both NOMA schemes considered. For the ease of presentation, we will slightly abuse the related notations
without causing ambiguity.

\subsection{Asynchronous NOMA Scheme}
\begin{figure}[htbp]
\centering
\subfigure[Diagram of aNOMA transmissions, where $K=3$ users are considered and their link delays are $0$, $\frac{2}{5}T$, and $\frac{6}{5}T$, respectively.]{
\centering
\includegraphics[width=0.8\textwidth]{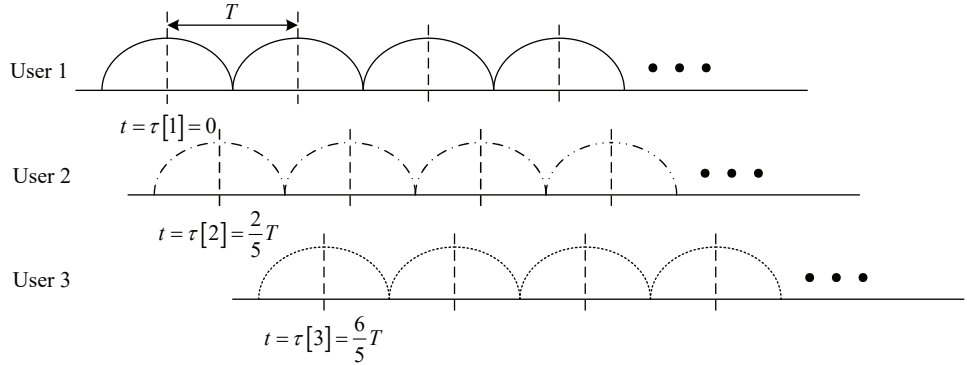}
\label{aNOMA_transmission}
}%
\quad
\subfigure[Diagram of aFTN-NOMA transmissions, where $K=3$ users are considered and their link delays are $0$, $\frac{2}{5}T$, and $\frac{6}{5}T$, respectively.]{
\centering
\includegraphics[width=0.8\textwidth]{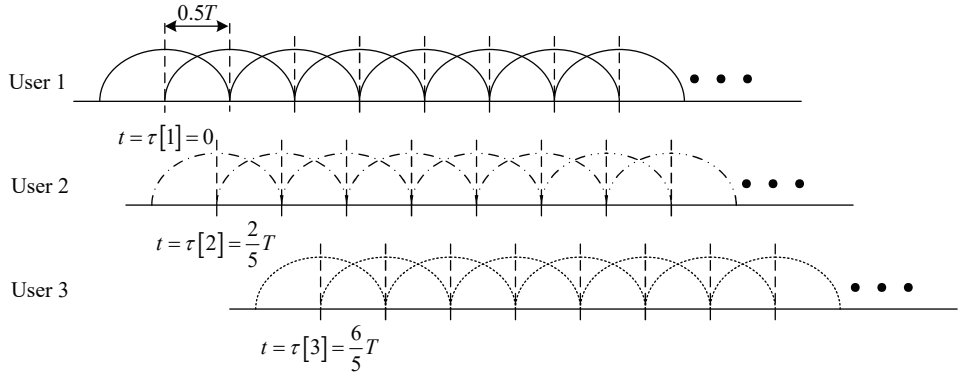}
\label{aFTN_NOMA_transmission}
}%
\quad
\subfigure[Diagram of NOMA transmissions, where $K=3$ users are considered.]{
\centering
\includegraphics[width=0.8\textwidth]{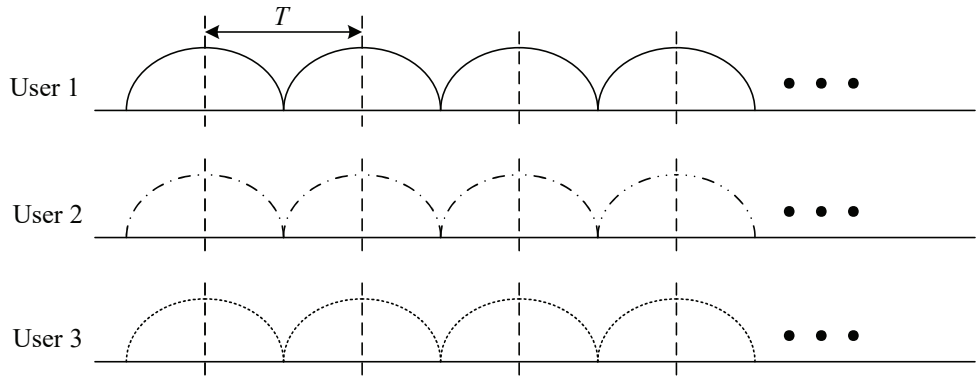}
\label{NOMA_transmission}
}%
\centering
\caption{%The uplink transmission diagram of considered scheme, where $K=3$ users are considered and their link delays are $0$, $\frac{2}{5}T$, and $\frac{6}{5}T$, respectively.
The uplink transmission diagram of considered scheme, where $K=3$ users are considered.}
\label{Diagrams}
%\vspace{-3mm}
\end{figure}

For the aNOMA scheme, the $k$-th user's transmitted signal is of the following form~\cite{ShuangyangWCNC}:
\begin{equation}
{s_k}\left( t \right) = \sqrt {{E_s}\left[ k \right]} \sum\limits_{n = 0}^{N - 1} {{x_k}\left[ n \right]p\left( {t - nT} \right)}  , \label{transmitted_signal_aNOMA}
\end{equation}
where ${{E_s}\left[ k \right]}$ is the average symbol energy of the $k$-th user.
Then, the signal $r\left( t \right)$ received by the BS is given by
\begin{equation}
r\left( t \right) = \sum\limits_{k = 1}^K {{h_k}{s_k}\left( {t - \tau \left[ k \right]} \right)}  + w\left( t \right) = \sum\limits_{k = 1}^K {\sum\limits_{n = 0}^{N - 1} {{h_k}\sqrt {{E_s}\left[ k \right]} {x_k}\left[ n \right]p\left( {t - nT - \tau \left[ k \right]} \right)} }  + w\left( t \right), \label{received_signal_aNOMA}
\end{equation}
where $w\left( t \right)\in {\mathbb C}$ is the additive white Gaussian noise (AWGN) at the BS with zero mean and one-sided PSD $N_0$.
For illustrating the asynchronous transmission, we provide a brief diagram in Fig.~\ref{aNOMA_transmission}, where we consider the uplink transmission of $3$ users having the specific link delays of $0$, $\frac{2}{5}T$, and $\frac{6}{5}T$, respectively.
%\begin{figure}
%\centering
%\includegraphics[width=0.6\textwidth]{Fig/aNOMA_transmission.eps}
%\caption{The uplink transmission diagram of aNOMA scheme, where $K=3$ users are considered and their link delays are $0$, $\frac{2}{5}T$, and $\frac{6}{5}T$, respectively.}
%\label{aNOMA_transmission}
%\vspace{-6mm}
%\centering
%\end{figure}

%\begin{figure}[htbp]
%  \centering
%  \begin{subfigure}{0.56\textwidth}
%    \includegraphics[width=0.6\textwidth]{Fig/aNOMA_transmission.eps}
%    \caption{First SUB-FLOAT}
%    \label{fig:demo1}
%  \end{subfigure}
%   \begin{subfigure}{0.56\textwidth}
%    \includegraphics[width=0.6\textwidth]{Fig/aFTN_NOMA_transmission.eps}
%    \caption{Second SUB-FLOAT}
%    \label{fig:demo2}
%  \end{subfigure}
%  \begin{subfigure}{0.56\textwidth}
%    \includegraphics[width=0.6\textwidth]{Fig/NOMA_transmission.eps}
%    \caption{Third SUB-FLOAT}
%    \label{fig:demo3}
%  \end{subfigure}
%  \caption{Demo}
%\end{figure}

Observe from the diagram that there is no ISI between the information symbols of a given user due to the $T$-orthogonal property, while each information symbol of the $k$-th user is interfered with all the information symbols of the other users due to the asynchronous transmission.
This is the ubiquitous multi-user interference (MUI).
By performing matched-filtering and Nyquist rate sampling for $r\left( t \right)$, the $n$-th element of the received symbol vector corresponding to the $k$-th user ${{\bf{y}}_k} = {\left[ {{y_k}\left[ 0 \right],{y_k}\left[ 1 \right], \ldots, {y_k}\left[ {N - 1} \right]} \right]^{\rm{T}}}$ is given by
\begin{align}
{y_k}\left[ n \right] \!&= \!\int\limits_{ - \infty }^\infty  \!\!{r\left( t \right){p^*}\left( {t - nT - \tau \left[ k \right]} \right)} {\rm{d}}t
 =\!\sum\limits_{l = 1}^K {\sum\limits_{m = 0}^{N - 1} {{h_l}\sqrt {{E_s}\left[ l \right]} {x_l}\left[ m \right]g\left[ {m \!- \! n,\tau \left[ l \right] \!-\! \tau \left[ k \right]} \right]} }  \!+\! {\eta _k}\left[ n \right]. \label{received_symbol_aNOMA}
\end{align}
In (\ref{received_symbol_aNOMA}), the term ${g\left[ {m - n,\tau \left[ l \right] - \tau \left[ k \right]} \right]}$ represents the MUI between different users for the aNOMA scheme, which is given by
\begin{equation}
g\left[ {\Delta k,{\Delta \tau  }} \right]  \buildrel \Delta \over =  \int\limits_{ - \infty }^\infty  {p\left( t \right){p^*}\left( {t + \Delta kT + {\Delta \tau  }} \right)} {\rm{d}}t
 =\int\limits_{ - \infty }^\infty  {{{\left| {{H_p}\left( f \right)} \right|}^2}\exp \left( {j2\pi f \left( {\Delta kT + {\Delta \tau }} \right)} \right)} {\rm{d}}f, \label{IUI_aNOMA}
\end{equation}
where the second equation is due to the Parseval's Theorem.
%According to~\eqref{IUI_aNOMA}, we have $g\left[ {k,0} \right] = 0$ for $1 - N \le k <0 $ and $0<k\le N - 1$ and $g\left[ {0,0} \right] = 1$, owing to the $T$-orthogonal property of the transmitter shaping pulse.
The term ${\eta _k}\left[ n \right]$ in~\eqref{received_symbol_aNOMA} denotes the corresponding colored noise sample, where $\mathbb{E}\left\{ {{\eta _k}\left[ n \right]\eta _l^*\left[ m \right]} \right\} = {N_0}g\left[ {m-n,\tau \left[ l \right] - \tau \left[ k \right]} \right]$.
For the ease of presentation,~\eqref{received_symbol_aNOMA} can be equivalently expressed in the matrix form of
\begin{equation}
{{\bf{y}}_k} = \sum\limits_{l = 1}^K {{h_l}\sqrt {{E_s}\left[ l \right]} {{\bf{G}}_{l,k}}{{\bf{x}}_l} + {{\bm{\eta }}_k}},  \label{vec_model_aNOMA}
\end{equation}
where ${{{\bf{G}}_{l,k}}}$ is the \emph{MUI channel matrix} of the aNOMA scheme, characterizing the MUI inflicted by the $l$-th user upon the $k$-th user, given by
\begin{equation}
{{\bf{G}}_{l,k}} = \left[ {\begin{array}{*{20}{c}}
{g\left[ {0,\tau \left[ l \right] - \tau \left[ k \right]} \right]}&{g\left[ {1,\tau \left[ l \right] - \tau \left[ k \right]} \right]}& \cdots &{g\left[ {N - 1,\tau \left[ l \right] - \tau \left[ k \right]} \right]}\\
{g\left[ { - 1,\tau \left[ l \right] - \tau \left[ k \right]} \right]}&{g\left[ {0,\tau \left[ l \right] - \tau \left[ k \right]} \right]}& \cdots &{g\left[ {N - 2,\tau \left[ l \right] - \tau \left[ k \right]} \right]}\\
 \vdots &{}& \ddots & \vdots \\
{g\left[ {1 - N,\tau \left[ l \right] - \tau \left[ k \right]} \right]}&{g\left[ {2 - N,\tau \left[ l \right] - \tau \left[ k \right]} \right]}& \cdots &{g\left[ {0,\tau \left[ l \right] - \tau \left[ k \right]} \right]}
\end{array}} \right].
\label{G_lk_aNOMA}
\end{equation}
It can be shown that ${\bf{G}}_{l,k}$ is a Toeplitz matrix and we have ${{\bf{G}}_{k,k}} = {{\bf{I}}_{N \times N}}$ for $1 \le k \le K$.
Meanwhile, the noise vector ${{\bm{\eta }}_k}$ is given by ${{\bm{\eta }}_k} = {\left[ {{\eta _k}\left[ 0 \right],{\eta _k}\left[ 1 \right],...,{\eta _k}\left[ {N - 1} \right]} \right]^{\rm{T}}}$, for $1 \le k \le K$, and it can be shown that $\mathbb{E}\left\{ {{\eta _k}\left[ m \right]\eta _k^*\left[ n \right]} \right\} = {N_0}$, for $1 \le k \le K$ and $0 \le n,m \le N - 1$.
\subsection{Asynchronous FTN-NOMA Scheme}
For the aFTN-NOMA scheme, the $k$-th user's transmitted signal is given by
\begin{equation}
{\tilde s_k}\left( t \right) = \sqrt {{E_s}\left[ k \right]} \sum\limits_{n = 0}^{N - 1} {{x_k}\left[ n \right]p\left( {t - n\zeta T} \right)}   , \label{transmitted_signal_aFTN_NOMA}
\end{equation}
where $\zeta  \in \left[ {0,1} \right]$ denotes the TD compression factor and the symbol period of the aFTN-NOMA scheme is $\zeta T$~\cite{li2017reduced,li2020time}.
Similar to the previous subsection, the signal ${\tilde r}\left( t \right)$ received by the BS is given by
\begin{equation}
\tilde r\left( t \right) = \sum\limits_{k = 1}^K {{h_k}{{\tilde s}_k}\left( {t - \tau \left[ k \right]} \right)}  + w\left( t \right) = \sum\limits_{k = 1}^K {\sum\limits_{n = 0}^{N - 1} {{h_k}\sqrt {{E_s}\left[ k \right]} {x_k}\left[ n \right]p\left( {t - n\zeta T - \tau \left[ k \right]} \right)} }  + w\left( t \right). \label{received_signal_aFTN_NOMA}
\end{equation}
%\begin{figure}
%\centering
%\includegraphics[width=0.6\textwidth]{Fig/aFTN_NOMA_transmission.eps}
%\caption{The uplink transmission diagram of aFTN-NOMA scheme, where $K=3$ users are considered with $\zeta=0.5$ and their link delays are $0$, $\frac{2}{5}T$, and $\frac{6}{5}T$, respectively.}
%\label{aFTN_NOMA_transmission}
%\vspace{-6mm}
%\centering
%\end{figure}

The transmission diagram of the aFTN-NOMA scheme having $\zeta=0.5$ is shown in Fig.~\ref{aFTN_NOMA_transmission}, where $3$ users having specific link delays $0$, $\frac{2}{5}T$, and $\frac{6}{5}T$ are considered.
In contrast to Fig.~\ref{aNOMA_transmission}, each information symbol is interfered by all the other information symbols among all users owing to both the FTN symbol rate and the asynchronous transmission.
For matched-filtering and FTN-rate sampling, the $n$-th element of the received symbol vector corresponding to the $k$-th user ${{\bf{y}}_k} = {\left[ {{y_k}\left[ 0 \right],{y_k}\left[ 1 \right], \ldots, {y_k}\left[ {N - 1} \right]} \right]^{\rm{T}}}$ is given by
\begin{align}
{y_k}\left[ n \right] &= \int\limits_{ - \infty }^\infty  {{\tilde r}\left( t \right){p^*}\left( {t - n\zeta T - \tau \left[ k \right]} \right)} {\rm{d}}t  \notag\\
 &=\sum\limits_{l = 1}^K {\sum\limits_{m = 0}^{N - 1} {{h_l}\sqrt {{E_s}\left[ l \right]} {x_l}\left[ m \right]{\tilde g}_{\zeta}\left[ {m - n,\tau \left[ l \right] - \tau \left[ k \right]} \right]} }  + {{\tilde \eta} _k}\left[ n \right]. \label{received_symbol_aFTN_NOMA}
\end{align}
In~\eqref{received_symbol_aFTN_NOMA}, the term ${\tilde g}_{\zeta}\left[ {\Delta k,{\Delta \tau  }} \right]$ represents the MUI between different information symbols given by
\begin{equation}
{\tilde g}_{\zeta}\left[ {\Delta k,{\Delta \tau  }} \right]  \buildrel \Delta \over = \!\!\! \int\limits_{ - \infty }^\infty  {p\left( t \right){p^*}\left( {t + \Delta k{\zeta}T + {\Delta \tau  }} \right)} {\rm{d}}t
\! =\!\!\!\int\limits_{ - \infty }^\infty  {{{\left| {{H_p}\left( f \right)} \right|}^2}\exp \left( {j2\pi f \left( {\Delta k{\zeta}T + {\Delta \tau }} \right)} \right)} {\rm{d}}f . \label{IUI_aFTN_NOMA}
\end{equation}
The term ${{\tilde {\eta}} _k}\left[ n \right]$ in~\eqref{received_symbol_aFTN_NOMA} denotes the corresponding colored noise sample, where $\mathbb{E}\left\{ {{{\tilde {\eta}} _k}\left[ n \right]{\tilde {\eta}}_l^*\left[ m \right]} \right\} = {N_0}{\tilde g}_{\zeta}\left[ {m-n,\tau \left[ l \right] - \tau \left[ k \right]} \right]$.
Similar to the previous subsection, we consider the equivalent matrix expression of~\eqref{received_symbol_aFTN_NOMA}, i.e.,
\begin{equation}
{{\bf{y}}_k} = \sum\limits_{l = 1}^K {{h_l}\sqrt {{E_s}\left[ l \right]} {\tilde{\bf{G}}_{l,k}}{{\bf{x}}_l} + {\tilde{\bm{\eta }}_k}},  \label{vec_model_aFTN_NOMA}
\end{equation}
where ${\tilde{\bf{G}}_{l,k}}$ is the \emph{MUI channel matrix} for the aFTN-NOMA scheme, characterizing the interference inflicted by the $l$-th user on the $k$-th user, given by
\begin{equation}
{\tilde{\bf{G}}_{l,k}} = \left[ {\begin{array}{*{20}{c}}
{{\tilde g}_{\zeta}\left[ {0,\tau \left[ l \right] - \tau \left[ k \right]} \right]}&{{\tilde g}_{\zeta}\left[ {1,\tau \left[ l \right] - \tau \left[ k \right]} \right]}& \cdots &{{\tilde g}_{\zeta}\left[ {N - 1,\tau \left[ l \right] - \tau \left[ k \right]} \right]}\\
{{\tilde g}_{\zeta}\left[ { - 1,\tau \left[ l \right] - \tau \left[ k \right]} \right]}&{{\tilde g}_{\zeta}\left[ {0,\tau \left[ l \right] - \tau \left[ k \right]} \right]}& \cdots &{{\tilde g}_{\zeta}\left[ {N - 2,\tau \left[ l \right] - \tau \left[ k \right]} \right]}\\
 \vdots &{}& \ddots & \vdots \\
{{\tilde g}_{\zeta}\left[ {1 - N,\tau \left[ l \right] - \tau \left[ k \right]} \right]}&{{\tilde g}_{\zeta}\left[ {2 - N,\tau \left[ l \right] - \tau \left[ k \right]} \right]}& \cdots &{{\tilde g}_{\zeta}\left[ {0,\tau \left[ l \right] - \tau \left[ k \right]} \right]}
\end{array}} \right].
\label{G_lk_aFTN_NOMA}
\end{equation}
Again, ${\tilde{\bf{G}}_{l,k}}$ is a Toeplitz matrix and the noise vector ${\tilde{\bm{\eta }}}_k$ is given by ${\tilde{\bm{\eta }}}_k= {\left[ {{{\tilde \eta} _k}\left[ 0 \right],{{\tilde \eta} _k}\left[ 1 \right],...,{{\tilde \eta} _k}\left[ {N - 1} \right]} \right]^{\rm{T}}}$, where ${\mathbb E}\left\{ {{{\tilde{\bm \eta}} _k}{\tilde{\bm \eta}} _k^H} \right\} = {N_0}{\widetilde {\bf{G}}_{k,k}}$.
\subsection{Connections to the Conventional Synchronous NOMA Scheme}
Conventionally, the impact of link delay difference is assumed to be perfectly eliminated by adaptive time frame alignment schemes
at the BS~\cite{WeiPerformanceGain,Zhiqiang2017optimal} for synchronous NOMA transmission.
Consequently, the signals transmitted from different users are perfectly aligned with each other at the BS, as shown in Fig.~\ref{NOMA_transmission}.
%\begin{figure}
%\centering
%\includegraphics[width=0.6\textwidth]{Fig/NOMA_transmission.eps}
%\caption{The uplink transmission diagram of conventional NOMA scheme, where $K=3$ users are considered.}
%\label{NOMA_transmission}
%\vspace{-5mm}
%\centering
%\end{figure}
By comparing aNOMA, aFTN-NOMA, and conventional synchronous NOMA schemes, it is plausible that the aFTN-NOMA scheme is the most general scheme.
More specifically, when the compression factor is $\zeta=1$, the aFTN-NOMA scheme degenerates to the aNOMA scheme. When there is no link delay difference among the users, the aNOMA scheme degenerates to the conventional synchronous NOMA scheme.

On the other hand, we notice that both the aNOMA and aFTN-NOMA schemes have at most $KN$ received symbols at the BS, while the conventional synchronous NOMA only has $N$ received symbols.
Furthermore, we also notice that the aNOMA, aFTN-NOMA and conventional synchronous NOMA schemes generally occupy different time resources.
Let ${\tau _{\max }} = \max \left\{ {\tau [1],\tau [2],...,\tau [K]} \right\}$ be the maximum link delay, which is usually negligible compared to the frame duration of the signals transmitted in practical systems~\cite{Zhiqiang2017optimal,Zhiqiang2019NOMAHybrid}.
To support each user transmitting $N$ information symbols, the conventional synchronous NOMA scheme roughly requires $NT$ seconds for its transmission, aNOMA scheme requires $NT+\tau_{\max}$ seconds, while the aFTN-NOMA scheme only requires $N \zeta T+\tau_{\max}$~\cite{Yuan2020Joint,Yuan2020iterative_SCMA}.

In the following, we investigate the achievable rates of both the aNOMA and aFTN-NOMA schemes based on~\eqref{vec_model_aNOMA} and~\eqref{vec_model_aFTN_NOMA}.
\section{Achievable Rate Analysis}
In this section, we focus our attention on the achievable rates of both the aNOMA and aFTN-NOMA schemes. We will first derive the closed-form expression of the mutual information for the uplink transmission and then apply Szeg\"o's Theorem~\cite{gray2006toeplitz,simon2010szegHo} to obtain further important insights. Particularly, Szeg\"o's Theorem is closely related to the DTFT of the underlying Toeplitz coefficients. Therefore, we will also investigate the characteristics of the DTFT with respect to the link delay and symbol rate.
For the ease of derivation, we assume that the elements in the transmitted symbol vector ${\bf{x}}_k$ are independent and identically distributed (i.i.d.) complex Gaussian variables with average symbol energy $E_s[k]$, for $\forall k$, $1\le k \le K$.
Since the aNOMA scheme can be viewed as a special case of the aFTN-NOMA scheme with $\zeta=1$, we will commence with the analysis of the aFTN-NOMA scheme.

Without loss of generality, let us assume that the channel coefficients are sorted in descending order, i.e., $|h_1|^2 \ge |h_2|^2 \ge... \ge |h_K|^2$.
Conventionally, SIC detection is applied at the BS for general NOMA systems{\footnote{We note that
some lattice-coding-based approaches can be applied to replace the SIC detection~\cite{qiu2019downlink,qiu2018lattice}.}}~\cite{WeiPerformanceGain}.
To analyze the achievable rates, it is commonly assumed that the MUI introduced by users $1,2,\ldots,k-1$ is perfectly cancelled for the detection of the $k$-th user~\cite{WeiPerformanceGain}.
Therefore, the asymptotic instantaneous achievable rate for the $k$-th user under SIC detection is given by
\begin{equation}
R_{{\bf{h}},{\bm{\tau}},\zeta}^{{k}} \buildrel \Delta \over = \mathop {\lim }\limits_{N \to \infty } \frac{1}{N}{{I_{{\bf{h}},{\bm{\tau}}, \zeta}}}\left( {{{\bf{y}}_k};{{\bf{x}}_k}|{{\bf{x}}_1}, \ldots ,{{\bf{x}}_{k - 1}}} \right) \quad {\rm bits \; per} \; {\rm{channel \; use}} .\label{achievable_rate}
\end{equation}
In particular, the closed-form expression of ${{I_{{\bf{h}},{\bm{\tau}},\zeta}}}\left( {{{\bf{y}}_k};{{\bf{x}}_k}| {{\bf{x}}_1}, \ldots ,{{\bf{x}}_{k-1}}} \right)$ is formulated in the following lemma.

\textbf{Lemma 1} (\emph{Conditional Mutual Information for SIC Detection}):
For SIC detection, the conditional mutual information ${{I_{{\bf{h}},{\bm{\tau}}, \zeta}}}\left( {{{\bf{y}}_k};{{\bf{x}}_k}| {{\bf{x}}_1}, \ldots ,{{\bf{x}}_{k-1}}} \right)$ of the aFTN-NOMA scheme is given by
\begin{align}
&{{I_{{\bf{h}},{\bm{\tau}},\zeta}}}\left( {{{\bf{y}}_k};{{\bf{x}}_k}| {{\bf{x}}_1}, \ldots ,{{\bf{x}}_{k-1}}} \right)\notag\\
=&\frac{1}{2}\log_2 \det \left[ {{{\bf{I}}_{N \times N}} + \frac{{{{\left| {{h_k}} \right|}^2}{E_s}\left[ k \right]{{{\bf{\tilde G}}}_{k,k}}{\bf{\tilde G}}_{k,k}^{\rm{T}}}}{{{N_0}}}{{\left( {{{{\bf{\tilde G}}}_{k,k}} + \frac{{\sum\limits_{l = k + 1}^K {{{\left| {{h_l}} \right|}^2}{E_s}\left[ l \right]} }}{{{N_0}}}{{{\bf{\tilde G}}}_{l,k}}{\bf{\tilde G}}_{l,k}^{\rm{T}}} \right)}^{ - 1}}} \right].  \label{Conditional_mutual_info_SIC}
\end{align}

\emph{Proof}: The proof is given in Appendix A.

The above equation is essentially the mutual information calculation over the channel with colored Gaussian noise, where the covariance matrix of the noise samples plus the interference is given by ${{N_0}{{{\bf{\tilde G}}}_{k,k}} + \sum\limits_{l = k + 1}^K {{{\left| {{h_l}} \right|}^2}{E_s}\left[ l \right]} {{{\bf{\tilde G}}}_{l,k}}{\bf{\tilde G}}_{l,k}^{\rm{T}}}$.
It can be observed from Lemma 1 that due to the link delay and the symbol rate of each user, the corresponding interference term of the aFTN-NOMA systems for the $k$-th user is different from that of the conventional synchronous NOMA systems.
With the help of Lemma 1, we now proceed to analyze the asymptotic instantaneous achievable rate $R_{{\bf{h}},{\bm{\tau}},\zeta}^{{k}}$ by invoking Szeg\"o's Theorem in order to obtain further important insights.
For reference, Szeg\"o's Theorem is stated as follows.

\textbf{Lemma 2} (\emph{Szeg\"o's Theorem~\cite{gray2006toeplitz,simon2010szegHo}}):
Let ${\bf{V}}$ denote a size $N \times N$ positive definite Toeplitz matrix ${\bf{V}}$, i.e.,
\begin{equation}
{\bf{V}} = \left[ {\begin{array}{*{20}{c}}
{{v_0}}&{{v_1}}& \cdots &{{v_{N - 1}}}\\
{{v_{ - 1}}}&{{v_0}}& \cdots &{{v_{N - 2}}}\\
 \vdots &{}& \ddots & \vdots \\
{{v_{1 - N}}}&{{v_{2 - N}}}& \cdots &{{v_0}}
\end{array}} \right],
\end{equation}
whose eigenvalues are given by $\left\{ {{\lambda _0},{\lambda _1}, \ldots {\lambda _{N - 1}}} \right\}$. Then, for an arbitrary continuous function $f_c(\cdot)$, we have
\begin{equation}
\mathop {\lim }\limits_{N \to \infty } \frac{1}{N}\sum\limits_{n = 0}^{N - 1} {{f_c}\left( {{\lambda _n}} \right) = \frac{1}{{2\pi }}\int_{ - \pi }^\pi  {{f_c}\left( {V\left( \omega  \right)} \right){\rm{d}}\omega } }  ,
\end{equation}
where ${V\left( \omega \right)}$ is the corresponding DTFT of the Toeplitz coefficients $\left\{ { \ldots ,{v_{ - 2}},{v_{ - 1}},{v_0},{v_1},{v_2}, \ldots } \right\}$, and it is given by
\begin{equation}
V\left( \omega  \right) = \sum\limits_{k =  - \infty }^\infty  {{v_k}{e^{ - jk\omega }}} .
\end{equation}

Szeg\"o's Theorem is a classic tool eminently suitable for analyzing the determinant of Toeplitz matrices, which is rooted in the FD characteristics of the Toeplitz coefficients, i.e., the DTFT. In particular, DTFT analysis has been widely applied in the research of signal sampling, which describes the FD representation of the underlying samples.
Let us define ${{\bf{T}}_{l,k}} \buildrel \Delta \over = {{{\bf{\tilde G}}}_{l,k}}{\bf{\tilde G}}_{l,k}^{\rm{T}}$ and
\begin{equation}
{{\bf{P}}_{k}} \buildrel \Delta \over = {{\bf{I}}_{N \times N}} + \frac{{{{\left| {{h_k}} \right|}^2}{E_s}\left[ k \right]{{\bf{T}}_{k,k}}}}{{{N_0}}}{\left( {{{{\bf{\tilde G}}}_{k,k}} + \frac{{\sum\limits_{l = k + 1}^K {{{\left| {{h_l}} \right|}^2}{E_s}\left[ l \right]} }}{{{N_0}}}{{\bf{T}}_{l,k}}} \right)^{ - 1}}.
\end{equation}
To apply Szeg\"o's Theorem, we first have to verify that ${{\bf{P}}_{k}}$ is a positive definite Toeplitz matrix in the asymptotic regime, i.e., $N \to \infty$, for any $1 \le k \le K$. Specifically, we have the following lemma.

\textbf{Lemma 3} (\emph{Positive Definiteness of the Asymptotical Toeplitz Matrix}):
As $N \to \infty$, both ${{{\bf{\tilde G}}}_{l,k}}$ and ${{\bf{T}}_{l,k}}$ are asymptotically positive definite Toeplitz matrices for any $1 \le k \le K$. Furthermore, ${{\bf{P}}_{k}}$ is also an asymptotically positive definite Toeplitz matrix for $1 \le k \le K$ as $N \to \infty$.

\emph{Proof}: The proof is given in Appendix B.

Next, we apply Szeg\"o's Theorem to \eqref{Conditional_mutual_info_SIC}. The application of Szeg\"o's Theorem requires the derivation of the DTFT of the Toeplitz coefficients.
As shown in~\eqref{Conditional_mutual_info_SIC}, there are two types of Toeplitz matrices, namely, ${{{\bf{\tilde G}}}_{l,k}}$ and ${{\bf{T}}_{l,k}}$.
In particular, the Toeplitz coefficients of ${{{\bf{\tilde G}}}_{l,k}}$ are given by $\left\{ {{{\tilde g}_\zeta }\left[ {n,\tau \left[ l \right] - \tau \left[ k \right]} \right]} \right\}$.
On the other hand, it can be shown that in the asymptotic regime, the Toeplitz coefficients $\left\{ {{t_{l,k}}\left[ n \right]} \right\}$ of the asymptotical Toeplitz matrix ${{\bf{T}}_{l,k}}$ are given by
\begin{equation}
{t_{l,k}}\left[ n \right] = \sum\limits_{m =  - \infty }^\infty  {{\tilde g}_{\zeta}\left[ {m,\tau \left[ l \right] - \tau \left[ k \right]} \right]{\tilde g}_{\zeta}\left[ {m - n,\tau \left[ l \right] - \tau \left[ k \right]} \right]} .
\label{t_l_k}
\end{equation}
Correspondingly, the DTFTs of the Toeplitz coefficients $\left\{ {{{\tilde g}_\zeta }\left[ {n,\tau \left[ l \right] - \tau \left[ k \right]} \right]} \right\}$ and $\left\{ {{t_{l,k}}\left[ n \right]} \right\}$ are given by
\begin{equation}
{{\tilde G}_{l,k}}\left( {2\pi f\zeta T} \right) = \sum\limits_{n =  - \infty }^\infty  {{{\tilde g}_\zeta }\left[ {n,\tau \left[ l \right] - \tau \left[ k \right]} \right]{e^{ - j2\pi n\zeta Tf}}} , \label{DTFT_G_l_k}
\end{equation}
and
\begin{equation}
{{\tilde T}_{l,k}}\left( {2\pi f\zeta T} \right) = \sum\limits_{n =  - \infty }^\infty  {\sum\limits_{m =  - \infty }^\infty  {{\tilde g}_\zeta\left[ {m,\tau \left[ l \right] - \tau \left[ k \right]} \right]{\tilde g}_\zeta\left[ {m - n,\tau \left[ l \right] - \tau \left[ k \right]} \right]} } {e^{ - j2\pi n\zeta Tf}}, \label{DTFT_T_l_k}
\end{equation}
respectively, where $\omega  \buildrel \Delta \over = 2\pi f\zeta T$.
We note that~\eqref{DTFT_G_l_k} and~\eqref{DTFT_T_l_k} depend on both the difference between the link delays of each user and the compression factor $\zeta$.
However, the link delays' difference is time-variant and it is therefore generally intractable.
As an alternative, we apply bounding techniques to the DTFTs to facilitate the achievable rate analysis. In particular, the bounds of DTFT depend on the corresponding spectra of the signaling pulse adopted.
Specifically, the spectrum of the RRC pulse with a roll-off factor $0 \le \beta \le 1$ is given by
\begin{equation}
{\left| {{H_p}\left( f \right)} \right|^2} = \left\{ \begin{array}{l}
T,\quad\quad\quad\quad\quad\quad\quad\quad\quad\quad\quad\left| f \right| < \left( {1 - \beta } \right)/\left( {2T} \right),\\
T{\cos ^2}\left( {\frac{{\pi T}}{{2\beta }}\left( {\left| f \right| - \frac{{1 - \beta }}{{2T}}} \right)} \right),\quad\left( {1 - \beta } \right)/\left( {2T} \right) \le \left| f \right| \le \left( {1 + \beta } \right)/\left( {2T} \right),\\
0,\quad\quad\quad\quad\quad\quad\quad\quad\quad\quad\quad\;\left| f \right| > \left( {1 + \beta } \right)/\left( {2T} \right).
\end{array} \right.\label{RRC_spectrum}
\end{equation}
As a building block for our rate analysis, we consider three FD signals related to ${\left| {{H_p}\left( f \right)} \right|^2}$ and the symbol rate $1/ \zeta T$.

\textbf{Definition 1} (\emph{Folded-Spectrum}):
Given the symbol rate $1/ \zeta T$ and the underlying signaling pulse spectrum ${\left| {{H_p}\left( f \right)} \right|^2}$, the \emph{folded-spectrum} is defined by
\begin{equation}
{\left| {{H_{{\rm{fo}}}}\left( f \right)} \right|^2} \buildrel \Delta \over = \sum\limits_{k =  - \infty }^\infty  {{{\left| {H_p\left( {f - \frac{k}{{\zeta T}}} \right)} \right|}^2}} , \label{folded_spectrum}
\end{equation}
for $f \in \left[ { - \frac{1}{{2 \zeta T}},\frac{1}{{2 \zeta T}}} \right]$ and zero otherwise.

\textbf{Definition 2} (\emph{Twisted Folded-Spectrum}):
Given the symbol rate $1/ \zeta T$ and the underlying signaling pulse spectrum ${\left| {{H_p}\left( f \right)} \right|^2}$, the \emph{twisted folded-spectrum} is defined by
\begin{equation}
{\left| {{{ H}_{{\rm{tfo}}}}\left( f \right)} \right|^2} \buildrel \Delta \over = {\left| {H_p\left( f \right)} \right|^2} - \sum\limits_{\scriptstyle k =  - \infty \hfill\atop
\scriptstyle k \ne 0\hfill}^\infty  {{{\left| {H_p\left( {f - \frac{k}{{\zeta T}}} \right)} \right|}^2}} , \label{twisted_folded_spectrum}
\end{equation}
for $f \in \left[ { - \frac{1}{{2 \zeta T}},\frac{1}{{2 \zeta T}}} \right]$ and zero otherwise.

\textbf{Definition 3} (\emph{Interference-Reducing-Spectrum}):
Given the symbol rate $\zeta T$ and the underlying signaling pulse spectrum ${\left| {{H_p}\left( f \right)} \right|^2}$, the \emph{interference-reducing-spectrum} is defined by
\begin{equation}
\rho \left( f \right) \buildrel \Delta \over = \frac{{{{\left| {{H_{{\rm{tfo}}}}\left( f \right)} \right|}^2}}}{{{{\left| {{H_{{\rm{fo}}}}\left( f \right)} \right|}^2}}} , \label{interference_reducing_spectrum}
\end{equation}
for $f \in \left[ { - \frac{1}{{2 \zeta T}},\frac{1}{{2 \zeta T}}} \right]$ and zero otherwise.

\begin{figure}[htbp]
\centering
\subfigure[${\left| {{H_{{\rm{fo}}}}\left( f \right)} \right|^2}$ for $\beta=0.5$, $\zeta=1$ and $\beta=0.5$, $\zeta\le 2/3$.]{
\begin{minipage}[t]{0.5\textwidth}
\centering
\includegraphics[scale=0.5]{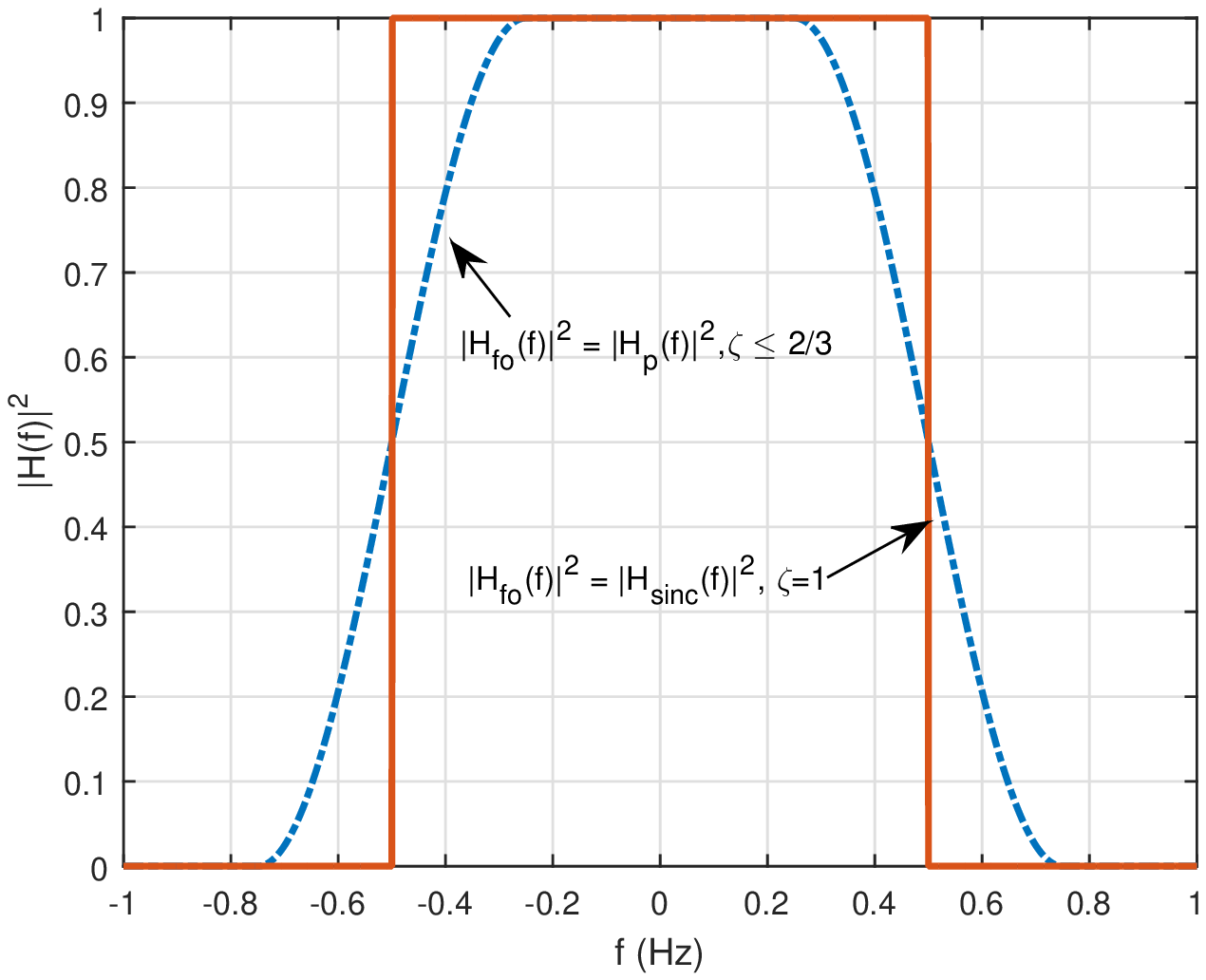}
%\caption{fig1}
\label{SpectrumsA}
\end{minipage}%
}%
\subfigure[${\left| {{H_{{\rm{tfo}}}}\left( f \right)} \right|^2}$ for $\beta=0.5$, $\zeta=1$ and $\beta=0.5$, $\zeta \le 2/3$.]{
\begin{minipage}[t]{0.5\textwidth}
\centering
\includegraphics[scale=0.5]{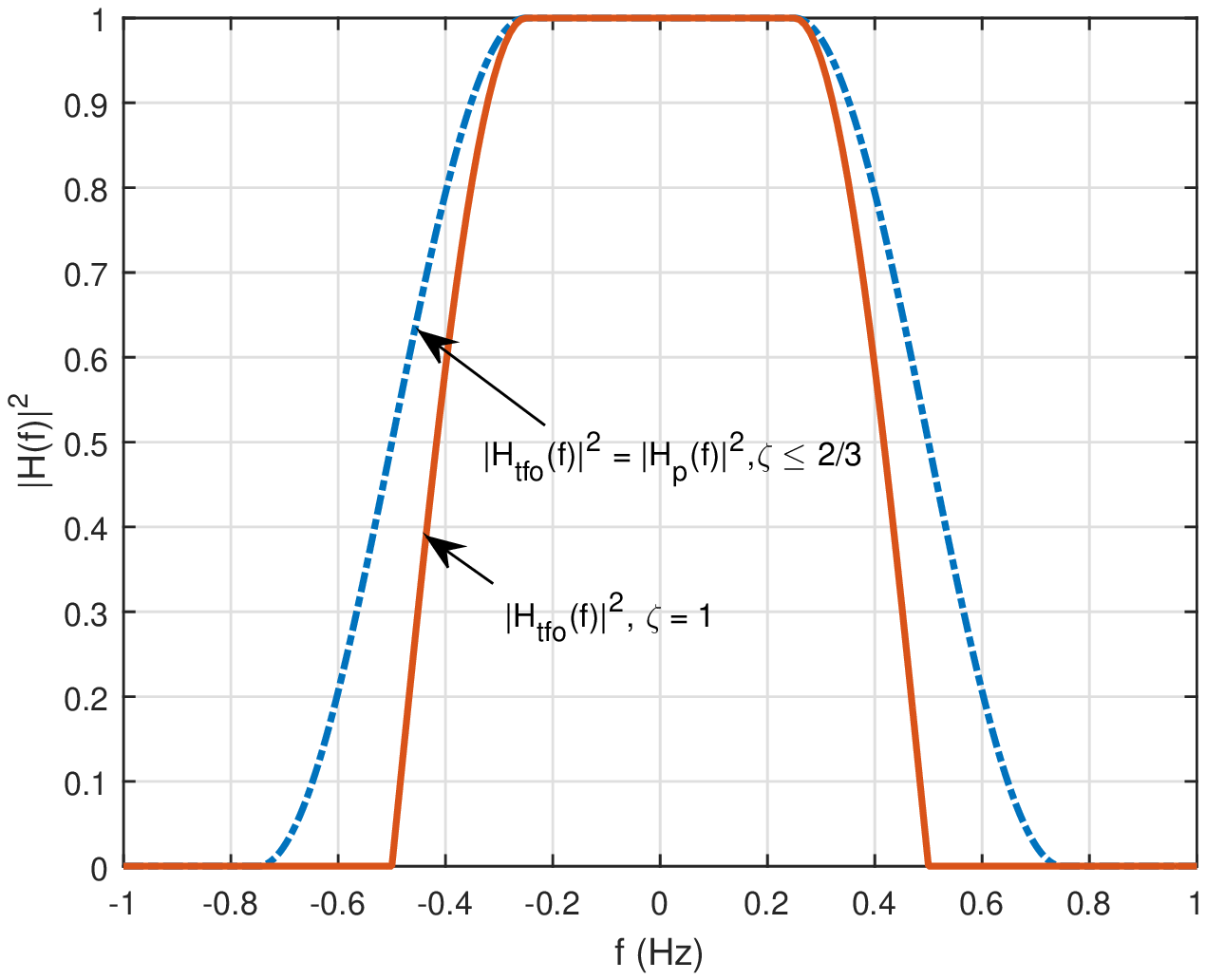}
%\caption{fig2}
\label{SpectrumsB}
\end{minipage}%
}%
\centering
\caption{Folded-spectrum and twisted folded-spectrum for $\beta=0.5$, $\zeta=1$ and $\beta=0.5$, $\zeta<2/3$, where $T=1$.}
\label{Spectrums}
\vspace{-3mm}
\end{figure}

Indeed, the folded-spectrum is commonly considered in the literature of faster-than-Nyquist signaling \cite{kim2016properties,kim2016faster} for the associated capacity analysis.
The folded-spectrum indicates that the frequency components outside the interval $\left[ { - \frac{1}{{2\zeta T}},\frac{1}{{2\zeta T}}} \right]$ are ``folded-in" the interval to form an equivalent FD representation of the transmitted signal. In particular, we refer to this folding effect as ``\textbf{spectral aliasing}". Spectral aliasing occurs, when the symbol rate is lower than twice the bandwidth of the signaling pulse, i.e., $\frac{1}{{2\zeta T}} < W$.
On the other hand, the twisted folded-spectrum can be viewed as a phase-rotated version of the folded-spectrum, which is useful for characterizing the FD phase rotation corresponding to the time delay.
Furthermore, the interference-reducing-spectrum is the ratio between the aforementioned two spectra, which can be viewed as
an indicator of how substantially the interference power is reduced due to the phase-rotation.
%Due to the symbol period $\zeta T$, we are interested in the shape of both folded-spectrum and twisted folded-spectrum in the frequency interval $f \in \left[ { - \frac{1}{{2 \zeta T}},\frac{1}{{2 \zeta T}}} \right]$.
Noticing that the RRC pulse is strictly bandlimited within the frequency interval of $f \in \left[ -W,W \right]$, it can be shown that ${\left| {{H_{{\rm{fo}}}}\left( f \right)} \right|^2}$ becomes the spectrum of the \emph{sinc} pulse ${\left| {H_{\rm sinc}\left( f \right)} \right|^2}$, i.e., the RRC pulse associated with $\beta=0$, when $\zeta=1$. On the other hand, both ${\left| {{H_{{\rm{fo}}}}\left( f \right)} \right|^2}$ and ${\left| {{{ H}_{{\rm{tfo}}}}\left( f \right)} \right|^2}$ become the exact RRC spectrum ${\left| {H_p\left( f \right)} \right|^2}$, when $\zeta \le 1/(1+\beta)$.
For reference, the plots of the folded-spectrum and twisted folded-spectrum for $\beta=0.5$, $\zeta=1$ and $\beta=0.5$, $\zeta<2/3$ are provided in Fig.~\ref{SpectrumsA} and Fig.~\ref{SpectrumsB}, respectively.
Furthermore, it may be readily seen that upon reducing $\zeta$, the value of the interference-reducing-spectrum $\rho \left( f \right)$ tends to $1$ for $f \in \left[ { - \frac{1}{{2 \zeta T}},\frac{1}{{2 \zeta T}}} \right]$. In particular, we have $\rho \left( f \right)=1$ for $f \in \left[ { - \frac{1}{{2 \zeta T}},\frac{1}{{2 \zeta T}}} \right]$, when $\zeta \le 1/(1+\beta)$.
This interesting fact actually indicates that when the symbol rate is sufficiently high, the impact of link delay tends to vanish as will be discussed in Section III-B.
We note that the folded-spectrum, the twisted folded-spectrum and the interference-reducing-spectrum are important for our analysis, because they are related to the signaling pulse, but they are affected differently with respected to the symbol rate $\frac{1}{\zeta T}$.
In the following lemma, we will unveil the intricate relationship between the inverse Fourier series corresponding to ${\left| {{H_p}\left( f \right)} \right|^2}$ and the aforementioned spectra.

\textbf{Lemma 4} (\emph{Bounds on the Infinite Fourier Series}):
Let $\gamma$ be an arbitrary constant number. Then, within the frequency interval $f \in \left[ { - \frac{1}{{2 \zeta T}},\frac{1}{{2\zeta T}}} \right]$, the infinite series  $\sum\limits_{k =  - \infty }^\infty  {{{\left| {{H_p}\left( {f - \frac{k}{\zeta T}} \right)} \right|}^2}} {e^{ - j2\pi \gamma k}}$ can be upper-bounded and lower-bounded by
\begin{equation}
{\left| {{H_{{\rm{tfo}}}}\left( f \right)} \right|^2} \le \sum\limits_{k =  - \infty }^\infty  {{{\left| {{H_p}\left( {f - \frac{k}{\zeta T}} \right)} \right|}^2}} {e^{ - j2\pi \gamma k}} \le {\left| {{H_{{\rm{fo}}}}\left( f \right)} \right|^2}, \label{Lemma4_bounds}
\end{equation}
%and can be approximated by
%\begin{equation}
%\sum\limits_{k =  - \infty }^\infty  {{{\left| {{H_p}\left( {f - \frac{k}{\zeta T}} \right)} \right|}^2}} {e^{ - j2\pi \gamma k}} \approx {\left| {{{ H}_p}\left( f \right)} \right|^2}, \label{Lemma4_approx}
%\end{equation}
%where the bounds in~\eqref{Lemma4_bounds} and the approximation in~\eqref{Lemma4_approx} become exact if $p\left( t \right)$ is the sinc pulse. Meanwhile, the upper-bound also becomes exact if $\gamma=0$.
where the bounds in~\eqref{Lemma4_bounds} become exact if $\beta=0$, i.e., $p\left( t \right)$ is the sinc pulse, or $\zeta \le \frac{1}{1+\beta}$. Meanwhile, the upper-bound also becomes exact if $\gamma=0$.

\emph{Proof}: The proof is given in Appendix C.

Lemma~4 characterizes the effect of phase-rotation on the folded-spectrum, showing that the phase-rotation may potentially change the shape of the spectrum.
Observe that the effect of spectral aliasing vanishes for $\beta=0$ or $\zeta \le \frac{1}{1+\beta}$. Therefore, Lemma~4 implies that the change due to the phase-rotation will
no longer exist, when there is no spectral aliasing. This is not unexpected, because when the spectra are sufficiently separated in the FD,
only the spectrum corresponding to $k=0$, i.e., ${{{\left| {{H_p}} (f)\right|}^2}}$, has non-zero values in the frequency interval $f \in \left[ { - \frac{1}{{2 \zeta T}},\frac{1}{{2\zeta T}}} \right]$. In this case, the infinite series $\sum\limits_{k =  - \infty }^\infty  {{{\left| {{H_p}\left( {f - \frac{k}{\zeta T}} \right)} \right|}^2}} {e^{ - j2\pi \gamma k}}$ reduces to ${\left| {{H_p}\left( f \right)} \right|^2}$, which is independent of the phase-rotation. Therefore, introducing phase-rotations (link delay) can no longer affect the value of the folded-spectrum, if there is no spectral aliasing.
Given Lemma 4, the DTFTs with respect to~\eqref{DTFT_G_l_k} and~\eqref{DTFT_T_l_k} can be upper- and lower-bounded for our analysis, as shown in the following theorems.

\textbf{Theorem 1} (\emph{Bounds on DTFT ${{\tilde G}_{l,k}}\left( {2\pi f\zeta T} \right)$}):
Given an arbitrary link delay difference $\Delta \tau  \buildrel \Delta \over = \tau \left[ l \right] - \tau \left[ k \right]$, the DTFT ${{\tilde G}_{l,k}}\left( {2\pi f\zeta T} \right)$ of the form~\eqref{DTFT_G_l_k} can be upper- and lower-bounded by
\begin{equation}
\frac{1}{{\zeta T}}{\left| {{H_{{\rm{tfo}}}}\left( f \right)} \right|^2}\le {{\tilde G}_{l,k}}\left( {2\pi f\zeta T} \right) \le\frac{1}{{\zeta T}} {\left| {{H_{{\rm{fo}}}}\left( f \right)} \right|^2}, \label{The_1}
\end{equation}
where the bounds in~\eqref{The_1} become exact if $\beta=0$, i.e., when $p\left( t \right)$ is the sinc pulse, or $\zeta \le \frac{1}{1+\beta}$. Meanwhile, the upper-bound also becomes exact if $\Delta \tau=0$, e.g., $l=k$.

\emph{Proof}: The proof is given in Appendix D.

\textbf{Theorem 2} (\emph{Bounds on DTFT ${{\tilde T}_{l,k}}\left( {2\pi f\zeta T} \right)$}):
Given an arbitrary link delay difference $\Delta \tau  \buildrel \Delta \over = \tau \left[ l \right] - \tau \left[ k \right]$, the DTFT ${{\tilde T}_{l,k}}\left( {2\pi f\zeta T} \right)$ of the form~\eqref{DTFT_T_l_k} can be upper- and lower-bounded by
\begin{equation}
{\left( {\frac{1}{{\zeta T}}{{\left| {{H_{{\rm{tfo}}}}\left( f \right)} \right|}^2}} \right)^2} \le {\tilde T_{l,k}}\left( {2\pi f\zeta T} \right) \le {\left( {\frac{1}{{\zeta T}}{{\left| {{H_{{\rm{fo}}}}\left( f \right)} \right|}^2}} \right)^2}, \label{The_2}
\end{equation}
where the bounds in~\eqref{The_1} become exact if $\beta=0$, i.e., if $p\left( t \right)$ is the sinc pulse, or $\zeta \le \frac{1}{1+\beta}$. Meanwhile, the upper-bound also becomes exact if $\Delta \tau=0$, e.g., $l=k$.

\emph{Proof}: The proof is given in Appendix E.

Based on the bounds on the DTFTs of the Toeplitz coefficients, we are ready to derive the bounds for the achievable rates. In particular, the main result is given in the following theorem.

\textbf{Theorem 3} (\emph{Bounds on the Achievable Rates of aFTN-NOMA Scheme}):
Let ${P_k} \buildrel \Delta \over = \frac{{{E_s}\left[ k \right]}}{{\zeta T}}$ denote the transmit power of the $k$-th user. Then, the asymptotic instantaneous achievable rate of aFTN-NOMA for the $k$-th user $R_{{\bf{h}},{\bm{\tau}},\zeta}^{{k}}$ under SIC detection is lower-bounded by
\begin{equation}
R_{{\bf{h}},{\bm{\tau}},\zeta}^{{k}} \ge \frac{1}{{2W}}\int_{ - \frac{1}{{2\zeta T}}}^{\frac{1}{{2\zeta T}}} {{{\log }_2}\left( {1 + \frac{{{{\left| {{h_k}} \right|}^2}{P_k}{{\left| {{H_{{\rm{fo}}}}\left( f \right)} \right|}^2}}}{{{N_0} + \sum\limits_{l = k + 1}^K  {{\left| {{h_l}} \right|}^2}{P_l}{{\left| {{H_{{\rm{fo}}}}\left( f \right)} \right|}^2}}}} \right)} {\rm{d}}f \quad {\rm{bits/s/Hz}}
\label{single_user_upper_bound_aFTN_NOMA}
\end{equation}
and upper-bounded by
\begin{align}
R_{{\bf{h}},{\bm{\tau}},\zeta}^{{k}} \le  \frac{1}{{2W}}\int_{ - \frac{1}{{2\zeta T}}}^{\frac{1}{{2\zeta T}}} {{{\log }_2}\left( {1 + \frac{{{{\left| {{h_k}} \right|}^2}{P_k}{{\left| {{H_{{\rm{fo}}}}\left( f \right)} \right|}^2}}}{{{N_0} + \sum\limits_{l = k + 1}^K  {{\left| {{h_l}} \right|}^2}{P_l}{{\left| {{H_{{\rm{tfo}}}}\left( f \right)} \right|}^2}\rho\left( f \right)}}} \right)} {\rm{d}}f \quad {\rm{bits/s/Hz}}.
\label{single_user_lower_bound_aFTN_NOMA}
\end{align}

\emph{Proof}: The proof is given in Appendix F.

With the help of Theorem~3, we now investigate the achievable rates of both the aNOMA and aFTN-NOMA schemes.
%For a better clarification, we will discuss the achievable rates of the two schemes separately in the following.

\subsection{Achievable Rates of Asynchronous NOMA Schemes}
For the aNOMA scheme, we have $\zeta=1$, where the folded-spectrum satisfies ${\left| {{H_{{\rm{fo}}}}\left( f \right)} \right|^2} = T$ within the frequency interval $f \in \left[ { - \frac{1}{{2T}},\frac{1}{{2T}}} \right]$.
Based on Theorem~3, the following corollary summarizes the bounds of the achievable rate for aNOMA schemes.

\textbf{Corollary 1} (\emph{Bounds on the Normalized Achievable Rates of aNOMA Scheme}):
The normalized asymptotic instantaneous achievable rate $R_{{\bf{h}},{\bm{\tau}},\zeta=1}^{{k}}$ of the aNOMA scheme for the $k$-th user under SIC detection is lower-bounded by
\begin{equation}
R_{{\bf{h}},{\bm{\tau}},\zeta=1}^{{k}} \ge \frac{1}{{2WT}} {{{\log }_2}\left( {1 + \frac{{{{\left| {{h_k}} \right|}^2}{P_k}T}}{{{N_0} + \sum\limits_{l = k + 1}^K {{{\left| {{h_l}} \right|}^2}} {P_l}T}}} \right)} \quad {\rm{bits/s/Hz}}
\label{single_user_lower_bound_aNOMA}
\end{equation}
and upper-bounded by
\begin{align}
R_{{\bf{h}},{\bm{\tau}},\zeta=1}^{{k}} \le \frac{1}{{2W}}\int_{ - \frac{1}{{2T}}}^{\frac{1}{{2T}}} {{{\log }_2}\left( {1 + \frac{{{{\left| {{h_k}} \right|}^2}{P_k}T}}{{{N_0} + \sum\limits_{l = k + 1}^K {{{\left| {{h_l}} \right|}^2}} {P_l}{{\left| {{H_{{\rm{tfo}}}}\left( f \right)} \right|}^2}\rho \left( f \right)}}} \right)} {\rm{d}}f \quad {\rm{bits/s/Hz}}.
\label{single_user_upper_bound_aNOMA}
\end{align}

\emph{Proof}: The corollary is a straightforward extension of Theorem~3 and thus the proof is omitted here.$\hfill\blacksquare$

According to Corollary~1, some interesting observations and insights can be revealed.
\begin{itemize}
\item Recalling Lemma~4, we observe that both the upper-bound and the lower-bound in Corollary~1 are achievable. Theoretically, the upper-bound can be achieved if the signaling pulse's FD roll-off factor is $\beta=0$, i.e., the sinc pulse, while the lower-bound can be achieved, if all the users share the same link delay, corresponding to the conventional synchronous NOMA system.
    %, and the corresponding achievable rate lower-bound is essentially the achievable rate for conventional synchronized NOMA schemes.

\item For practical RRC pulses, i.e., $\beta \ne 0$, the asynchronous transmission leads to an improved achievable rate region compared to that of the conventional synchronous NOMA systems.
    %In specific, as  ${{\left| {{{ H}_{{p}}}\left( f \right)} \right|}^4} \le T^2$ within the frequency interval $f \in \left[ { - \frac{1}{2T},  \frac{1}{2T}} \right]$, it is obvious that the aNOMA system enjoys a smaller multiuser interference caused by different link delays.
    Moreover, the upper-bound indicates that the potential data rate improvement of aNOMA systems is due to its reduced MUI energy, yielding an SINR improvement, which is the result of different link delays.
    An intuitive explanation of this observation is that the aNOMA system naturally avoids the full superposition of maximum MUI owing to the diverse link delays.
    Hence, aNOMA system is unlikely to suffer from the peak interference energy at each sampling instant, which is consistent with the observations in Fig.~\ref{aNOMA_transmission}.
    However, for the sinc pulse, i.e., $\beta = 0$, the asynchronous transmission does not provide any rate improvement. This is because the symbol rate is consistent with the bandwidth of the sinc pulse. Hence, no spectral aliasing occurs and thus the introduction of phase-rotations cannot improve the achievable rate.
    This observation indicates that compared to conventional synchronous NOMA systems, the aNOMA system suffers from less severe MUI caused by the different link delays and offers the potential of achieving higher rates.
    %On the other hand, compared with conventional OMA systems, aNOMA systems allow multiple users to transmit their information over almost the same time-frequency resources at a cost of manageable multiuser interference.
\item The instantaneous achievable rate region of the aNOMA system is directly determined by both the folded-spectrum and the twisted folded-spectrum. In particular, the
    achievable rate improvement due to the asynchronous transmission increases upon increasing $\beta$, since the corresponding twisted folded-spectrum has a low energy in the frequency interval $f \in \left[ { - \frac{1}{{2T}},\frac{1}{{2T}}} \right]$. However, compared to the zero excess bandwidth of $\beta=0$, the normalized achievable rates of both the aNOMA scheme and the conventional synchronous NOMA scheme are reduced upon increasing of $\beta$ due to the normalization.
\end{itemize}

%In summary, we note that the aNOMA scheme generally enjoys a larger achievable rate region compared to the conventional synchronized NOMA scheme with the same channel conditions. The correctness of the derived bounds will be verified in our numerical results.

\subsection{Achievable Rates of Asynchronous FTN-NOMA Schemes}
We have already formulated the achievable rates of the aFTN-NOMA schemes in Theorem~3.
In particular, for $\zeta<1$, we observe that the aFTN-NOMA scheme enjoys both an SINR gain granted by the asynchronous transmissions (corresponding to the interference-reducing-spectrum) and a DoF gain introduced by the FTN transmission (corresponding to the integral range).
To further explain advantages of aFTN-NOMA, let use formally define the SINR gain and DoF gain of aFTN-NOMA schemes as follows.
%In what follows, we will further discuss the achievable rates improvements of aFTN-NOMA schemes with respect to different compression factor $\zeta$.

\textbf{Definition 4} (\emph{SINR Gain over Synchronous Transmission}):
The SINR gain of aFTN-NOMA schemes over the synchronous schemes having the same symbol rate is defined by
\begin{equation}
{\rm{Gai}}{{\rm{n}}_{{\rm{SINR}}}} \buildrel \Delta \over = \frac{{\int_{ - \frac{1}{{2\zeta T}}}^{\frac{1}{{2\zeta T}}} {\frac{{{{\left| {{h_k}} \right|}^2}{P_k}{{\left| {{H_{{\rm{fo}}}}\left( f \right)} \right|}^2}}}{{{N_0} + \sum\limits_{l = k + 1}^K {{{\left| {{h_l}} \right|}^2}{P_l}{{\left| {{H_{{\rm{tfo}}}}\left( f \right)} \right|}^2}\rho \left( f \right)} }}} {\rm{d}}f}}{{\int_{ - \frac{1}{{2\zeta T}}}^{\frac{1}{{2\zeta T}}} {\frac{{{{\left| {{h_k}} \right|}^2}{P_k}{{\left| {{H_{{\rm{fo}}}}\left( f \right)} \right|}^2}}}{{{N_0} + \sum\limits_{l = k + 1}^K {{{\left| {{h_l}} \right|}^2}{P_l}{{\left| {{H_{{\rm{fo}}}}\left( f \right)} \right|}^2}} }}} {\rm{d}}f}}
=\int_{ - \frac{1}{{2\zeta T}}}^{\frac{1}{{2\zeta T}}} {\frac{{{N_0}{{\left| {{H_{{\rm{fo}}}}\left( f \right)} \right|}^2} + \sum\limits_{l = k + 1}^K {{{\left| {{h_l}} \right|}^2}{P_l}{{\left| {{H_{{\rm{fo}}}}\left( f \right)} \right|}^4}} }}{{{N_0}{{\left| {{H_{{\rm{fo}}}}\left( f \right)} \right|}^2} + \sum\limits_{l = k + 1}^K {{{\left| {{h_l}} \right|}^2}{P_l}{{\left| {{H_{{\rm{tfo}}}}\left( f \right)} \right|}^4}} }}} {\rm{d}}f. \label{SINR_gain}
\end{equation}

\textbf{Definition 5} (\emph{DoF Gain over Conventional NOMA}):
Let us define the \emph{effective bandwidth} ${\rm{B}}{{\rm{W}}_{{\rm{eff}}}}\left( {\frac{1}{{\zeta T}}} \right)$ with respect to the symbol rate ${\frac{1}{{\zeta T}}}$ as the frequency interval, where both the folded-spectrum and the twisted folded-spectrum have non-zero values, i.e., ${\rm{B}}{{\rm{W}}_{{\rm{eff}}}} \left( {\frac{1}{{\zeta T}}} \right)\buildrel \Delta \over = \min \left\{ {\frac{1}{{\zeta T}},2W} \right\}$.
Then, the DoF gain of the aFTN-NOMA scheme over the conventional NOMA scheme is defined by the ratio between the effective bandwidths of the two schemes, i.e.,
\begin{equation}
{\rm{Gai}}{{\rm{n}}_{{\rm{DoF}}}} \buildrel \Delta \over = \frac{{{\rm{B}}{{\rm{W}}_{{\rm{eff}}}}\left( {\frac{1}{{\zeta T}}} \right)}}{{{\rm{B}}{{\rm{W}}_{{\rm{eff}}}}\left( {\frac{1}{T}} \right)}} = T \times \min \left\{ {\frac{1}{{\zeta T}},2W} \right\}.\label{DoF_gain}
\end{equation}

Observe that with the reduction of $\zeta$, the SINR gain vanishes according to the properties of ${\left| {{H_{{\rm{fo}}}}\left( f \right)} \right|}^2$ and ${\left| {{H_{{\rm{tfo}}}}\left( f \right)} \right|}^2$, i.e.,
${\left| {{H_{{\rm{fo}}}}\left( f \right)} \right|}^2-{\left| {{H_{{\rm{tfo}}}}\left( f \right)} \right|}^2$ is a non-increasing function in the range of $0 \le \zeta \le 1$.
Furthermore, we can also observe that the DoF gain increases upon reducing $\zeta$.
Therefore, we can see that there exists an intriguing trade-off between the SINR gain and DoF gain of the aFTN-NOMA schemes with respect to the symbol rate (compression factor).

\textbf{Proposition 1} (\emph{Trade-off Between SINR Gain and DoF Gain}):
Upon increasing the symbol rate from $\zeta=1$ to $\zeta=1/(1+\beta)$, the SINR gain of the aFTN-NOMA scheme decreases, while the DoF gain increases, and \emph{vice versa}.

Particularly, it can be shown that the SINR gain achieves its maximum value when $\zeta=1$, i.e., for the aNOMA scheme, where there is no DoF gain.
Furthermore, the DoF gain achieves its maximum when $\zeta \le 1/(1+\beta)$, where there is no SINR gain. Specifically, Definition~1 and Definition~2 have shown that both the folded-spectrum and the twisted folded-spectrum become exactly the same as the RRC spectrum within the frequency interval of $f \in\left[ { - \frac{1}{{2\zeta T}},\frac{1}{{2\zeta T}}} \right]$ for $\zeta\le 1/(1+\beta)$, in which case the upper- and lower-bounds in Theorem~3 are merged together and the corresponding maximum DoF gain is of value $2WT$. %aligning with the Shannon's famous $2 \cal W \cal T$ Theorem~\cite{shannon1949communication}.
Consequently, $R_{{\bf{h}},{\bm{\tau}},\zeta\le 1/(1+\beta)}^{{k}}$ of the aFTN-NOMA schemes associated with $\zeta\le 1/(1+\beta)$ is given by
\begin{equation}
R_{{\bf{h}},{\bm{\tau}},\zeta \le \frac{1}{{1 + \beta }}}^{{k}} =\frac{1}{{2W}}\int_{ - W}^{W} {{{\log }_2}\left( {1 + \frac{{{{\left| {{h_k}} \right|}^2}{P_k}{{\left| {{H_p}\left( f \right)} \right|}^2}}}{{{N_0} + \sum\limits_{l = k + 1}^K {{{\left| {{h_l}} \right|}^2}} {P_l}{{\left| {{H_p}\left( f \right)} \right|}^2}}}} \right)} {\rm{d}}f \quad {\rm{bits/s/Hz}}.
\label{single_user_aFTN_NOMA_high_symbol_rate}
\end{equation}
Given~\eqref{single_user_aFTN_NOMA_high_symbol_rate}, we can characterize the rate improvement of the aFTN-NOMA scheme over the conventional synchronous NOMA scheme in the high-SNR regime by the following corollary.

\textbf{Corollary 3} (\emph{Achievable Rate Improvement of aFTN-NOMA Schemes}):
Given a sufficiently high SNR, the asymptotic instantaneous achievable rate of aFTN-NOMA schemes is $1+\beta$ times higher than that of the conventional synchronous NOMA scheme.

\emph{Proof}: Considering the asymptotic instantaneous achievable rate of the conventional synchronous NOMA scheme given in~\eqref{single_user_lower_bound_aNOMA}, we have
\begin{align}
&\mathop {\lim }\limits_{{N_0} \to 0} \frac{{\frac{T}{{1 + \beta }}\int_{ - \frac{{1 + \beta }}{{2T}}}^{\frac{{1 + \beta }}{{2T}}} {{{\log }_2}\left( {1 + \frac{{{{\left| {{h_k}} \right|}^2}{P_k}{{\left| {{H_p}\left( f \right)} \right|}^2}}}{{{N_0} + \sum\limits_{l = k + 1}^K {{{\left| {{h_l}} \right|}^2}} {P_l}{{\left| {{H_p}\left( f \right)} \right|}^2}}}} \right){\rm{d}}f} }}{{\frac{1}{{2WT}}}{{{\log }_2}\left( {1 + \frac{{{{\left| {{h_k}} \right|}^2}{P_k}T}}{{{N_0} + \sum\limits_{l = k + 1}^K {{{\left| {{h_l}} \right|}^2}} {P_l}T}}} \right)}}
= \frac{{{{\log }_2}\left( {1 + \frac{{{{\left| {{h_k}} \right|}^2}{P_k}}}{{\sum\limits_{l = k + 1}^K {{{\left| {{h_l}} \right|}^2}} {P_l}}}} \right)}}{{\frac{1}{{2WT}}{{\log }_2}\left( {1 + \frac{{{{\left| {{h_k}} \right|}^2}{P_k}}}{{\sum\limits_{l = k + 1}^K {{{\left| {{h_l}} \right|}^2}} {P_l}}}} \right)}}
={1 + \beta }.
\end{align}
This completes the proof of Corollary~3.

According to the conclusions from Theorem~3, Proposition 1 and Corollary~3, some important insights can be revealed for the family of aFTN-NOMA schemes.
\begin{itemize}
\item Similar to the aNOMA scheme, the achievable rate improvement of aFTN-NOMA schemes stems from the excess bandwidth of the signaling pulse, where both the upper-bound and the lower-bound in Theorem~3 are achievable according to Lemma~4. Given a sufficiently high symbol rate, the upper- and lower-bounds in Theorem~3 are merged together, as suggested in~\eqref{single_user_aFTN_NOMA_high_symbol_rate}.
\item Proposition~1 has demonstrated the trade-off between the SINR gain and DoF gain. The physical interpretation of this trade-off is as follows. Given a higher symbol rate, the effect of spectral aliasing is mitigated, in which case the change of the folded-spectrum due to the phase-rotation is limited. In particular, when $\zeta \le 1/(1+\beta)$, the spectral aliasing no longer exists and thus the SINR gain vanishes.
%\item Compared to the SINR gain, the DoF gain essentially leads to a much higher achievable rate in the high SNR regime. It particular, Corollary~3 has shown that the aFTN-NOMA scheme can transmit $1+\beta$ times more information compared with the conventional synchronized NOMA scheme with a sufficiently high SNR. In fact, the achievable rate in~\eqref{single_user_aFTN_NOMA_high_symbol_rate} is essentially the ultimate achievable rate with a given transmitter pulse and an SINR~\cite{rusek2009constrained}. This is because the symbol rate is no smaller than the bandwidth of the shaping pulse and therefore both the folded-spectrum and the twisted folded-spectrum become exactly the same of the underlying RRC spectrum, yielding the maximum capacity promised by the shaping pulse.
%    However, for the sinc pulse, i.e., $\beta = 0$, a higher symbol rate does not provide any improvement in terms of achievable rates, as the Nyquist symbol rate equals to twice of the bandwidth of the shaping pulse, i.e., no spectrum aliasing.
%    Different from the aNOMA scheme and the conventional synchronized NOMA scheme, the normalized achievable rates for the aFTN-NOMA scheme with a sufficiently high symbol rate only exhibits a constant gap to the $\beta=0$ case in the high SNR regime. This is because the FTN transmission can fully explore the DoF promised by the transmitter shaping pulse~\cite{rusek2009constrained}.
\item Compared to the SINR gain, the DoF gain essentially leads to a higher achievable rate in the high-SNR regime, as shown in Corollary~3. In fact, the rate in~\eqref{single_user_aFTN_NOMA_high_symbol_rate} is essentially the ultimate achievable rate for a given signaling pulse and an SINR~\cite{rusek2009constrained}. This is because the symbol rate is no lower than the bandwidth of the signaling pulse and therefore both the folded-spectrum and the twisted folded-spectrum become exactly the same as the underlying RRC spectrum, yielding the maximum capacity promised by the signaling pulse.
    However, for the sinc pulse, i.e., $\beta = 0$, a higher symbol rate does not provide any improvement in terms of achievable rates, since the Nyquist symbol rate is simply twice the bandwidth of the signaling pulse, i.e., there is no spectral aliasing.
    In contrast to the aNOMA scheme and to the conventional synchronous NOMA scheme, the normalized achievable rate of the aFTN-NOMA scheme attained at a sufficiently high symbol rate only exhibits a constant gap to with respect to the $\beta=0$ case in the high-SNR regime. This is because the FTN transmission successes in fully exploiting the DoF promised by the siganling pulse~\cite{rusek2009constrained}.
\end{itemize}

\subsection{Implementations, Comparisons, and Extensions of the Considered Schemes}
We have shown the advantages of aFTN-NOMA and aNOMA over the conventional synchronous NOMA scheme in terms of their achievable rate. However, it should be noted that both aFTN-NOMA and aNOMA suffer from ISI and therefore, they tend to impose an increased detection complexity~\cite{Yuan2020iterative}.
A promising detector design could be based on the combination of both the MUI and ISI detections. In particular, the MUI can be efficiently dealt with by the SIC technique, given the channel disparities~\cite{WeiPerformanceGain}, while the ISI arising from both FTN and asynchronous transmissions can be mitigated by various mature detection methods conceived for ISI channels, e.g.,~\cite{rusek2011optimal,ishihara2021evolution}. Although the related research on the detection issue is at its early stage, several potent detectors are available in the literature. We refer the interested readers to~\cite{Yuan2020iterative} and~\cite{Ameha2018iterative} for more details.

Note that our previous analysis in this section is purely based on baseband signaling, where we show that both aNOMA and aFTN-NOMA are superior to the conventional synchronous NOMA in terms of their achievable rates with arbitrarily given channel coefficients. Therefore, it is expected that both aFTN-NOMA and aNOMA will outperform conventional NOMA with any given distribution of channel coefficients. For a better understanding of aNOMA and aFTN-NOMA, we briefly compare them to similar MA schemes as follows.
\begin{itemize}
\item \textbf{aFTN-NOMA vs. OMA with FTN signaling}: According to the previous analysis in this section, the achievable rate improvements attained by asynchronous transmission and FTN signaling accrue from the excess bandwidth of the signaling pulse, which is independent from the gain of NOMA gleaned from exploiting power discrepancies among different users.
    Furthermore, NOMA allows the information from different users to be transmitted in a time-sharing manner, which can be shown to have a higher achievable rate region than OMA~\cite{cover2012elements}. Therefore, aFTN-NOMA generally has a better performance than OMA with FTN signaling in terms of achievable rates, because OMA cannot exploit the power discrepancies and does not allow time-sharing among different users.
\item   \textbf{aNOMA/aFTN-NOMA vs. RSMA}: The success of RSMA lies in the rate-splitting, where the message of each user is divided into two parts, namely, a common part and a private part~\cite{Clerckx2021critical}. As the common part is supposed to be decoded by all the users, RSMA generally enjoys a reduced MUI compared to the conventional power-domain NOMA~\cite{Clerckx2021critical}. Compared to RSMA, both aNOMA and aFTN-NOMA reduce the MUI by relying on asynchronous transmission, which is different from the principle of RSMA. However, it is generally not fair or practical to directly compare whether aNOMA/aFTN-NOMA or RSMA has inflicts lower MUI, because they both rely on the channel conditions, including the distributions of the channel coefficients and link delays. It is also worth mentioning that the advantages of both  aNOMA and aFTN-NOMA become more pronounced for a higher excess bandwidth of the shaping pulse, while RSMA cannot make use of the excess bandwidth. %Furthermore, aFTN-NOMA is able to obtain the DoF gain based on the excess bandwidth and therefore, aFTN-NOMA may outperform RSMA is the high SNR regime.
    An interesting discussion at this point may be the combination of asynchronous transmission and FTN signaling with RSMA. As RSMA and aNOMA/aFTN-NOMA enjoy advantages over NOMA from a range of different perspectives, their combination might lead to further rate improvements, which will be considered in our future work.
\end{itemize}

Now, we briefly discuss the potential extensions of the analysis in the previous subsections to more practical systems, including multi-carrier, multi-antenna, and multi-cell systems.
\begin{itemize}
\item \textbf{Extension to multi-carrier systems}: In multi-carrier systems, the signaling pulse is usually time-limited instead of being bandlimited. Thus, the schemes considered may be extended to multi-carrier systems by allowing asynchronous transmission and FTN signaling in the FD instead of the TD, which is essentially a type of spectrally efficient frequency domain multiplexing (FDM) signals~\cite{kanaras2009spectrally,darwazeh2013optical} with asynchronous transmissions. Our previous analysis could be extended to this case by interchanging the corresponding analysis between the FD and the TD.
\item \textbf{Extension to multi-antenna systems}: Assume that the BS is equipped with multiple receive antennas, while the user has only a single transmit antenna. In this case, the BS may receive multiple copies of the transmitted signal~\eqref{received_signal_aNOMA} or~\eqref{received_signal_aFTN_NOMA} of each user at different link delays and channel coefficients. Naturally, SIC detection could be applied at the receiver for multiuser signal detection. The achievable rate analysis of such a system may also rely on the Toeplitz structure of the corresponding channel matrix after suitable combining of the signals received from different antennas.
\item \textbf{Extension to multi-cell systems}: In multi-cell systems, the transmitted signals could be received by multiple cells. Then, depending on whether the cooperation between different cells is allowed, the inter-cell interference could be either exploited by cooperation or simply treated as noise for low-complexity processing. It should be noted that the signals transmitted by different users cannot be received by more than two BSs with perfect synchronization in practice, hence the considered asynchronous transmissions are permanently suitable for cooperative multi-cell systems. The achievable rate analysis of multi-cell systems may also rely on the analysis of
    single-cell settings. However, how the extra channel diversity gleaned from the potential multi-cell cooperations will improve the achievable rate may require further investigations.
\end{itemize}
Unfortunately, we have to leave the above interesting issues for our future works due to the page limitation.
\section{Numerical Results}
In this section, we compare the normalized achievable rates of the aNOMA scheme and the aFTN-NOMA scheme to that of the conventional synchronous NOMA scheme, where the actual achievable rate of the considered schemes are obtained based on~\eqref{Conditional_mutual_info_SIC} via the classic Monte Carlo method.
In particular, we compare both the instantaneous rates and the ergodic rates relying on the randomly generated link delay for each user, where the link delay is assumed to follow a
uniform distribution within the interval $[0, 2 T]$ for each Monte Carlo realization.
In order to verify our previous discussions, the instantaneous rates are calculated based on a typical NOMA transmission supporting $K=2$ and $K=3$ users, and the number of transmitted symbols for each user is set to $N=100$.
On the other hand, the ergodic rates are calculated based on a typical single-cell settings from~\cite{WeiPerformanceGain}.
Without loss of generality, we adopt the equal power allocation for the different users for all the related simulations, i.e., $P_k=P$, for $k=1,2,...K$.

\subsection{Normalized Instantaneous Achievable Rates for the Two-User Case}
We consider a specific channel realization, where the channel coefficients for the two users are given by  ${{\left| {{h_1}} \right|}^2}=0.5$ and ${{\left| {{h_2}} \right|}^2}=0.5$ and we have $\frac{{ {{{\left| {{h_1}} \right|}^2}{P_1}} }}{{{N_0}}}=\frac{{ {{{\left| {{h_2}} \right|}^2}{P_2}} }}{{{N_0}}} = 10$ dB. In order to obtain a general conclusion, we assume that each user has a random link delay. Given the channel coefficients, we adopt the Monte Carlo method to average the achievable rate with different link delays.

\begin{figure}
\centering
\includegraphics[width=0.6\textwidth]{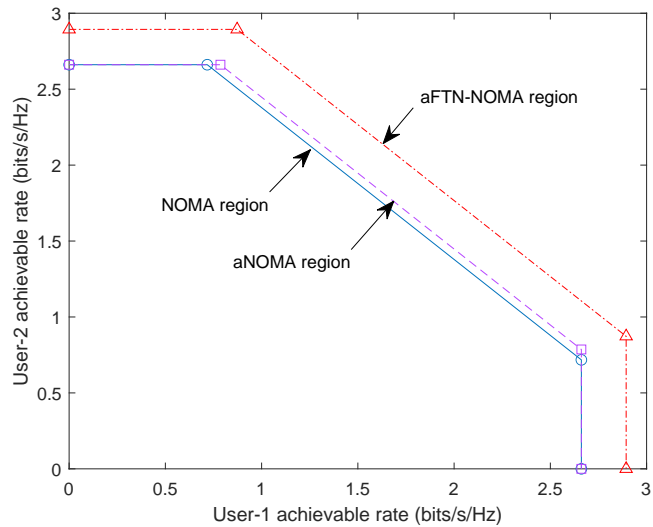}
\caption{The achievable rate regions of the conventional synchronous NOMA, the aNOMA, and the aFTN-NOMA schemes for $K=2$ users, where $\frac{{ {{{\left| {{h_1}} \right|}^2}{P_1}} }}{{{N_0}}}=\frac{{ {{{\left| {{h_2}} \right|}^2}{P_2}} }}{{{N_0}}} = 10$ dB. The signaling pulse is the RRC pulse with $\beta=0.3$.}
\label{Two_User}
\centering
\vspace{-6mm}
\end{figure}

The achievable rate regions of the conventional synchronous NOMA, the aNOMA, and the aFTN-NOMA schemes are compared in Fig.~\ref{Two_User}, where the signaling pulse is the RRC pulse using $\beta=0.3$ and the compression factor for the aFTN-NOMA scheme is $\zeta=0.75$.
As shown in the figure, the conventional synchronous NOMA has the smallest achievable rate region among all the three schemes, while the aNOMA scheme only shows a marginal improvement. On the other hand, the aFTN-NOMA scheme can considerably improve the achievable rate as shown in Fig.~\ref{Two_User}, which is consistent with our previous analysis.

\subsection{Normalized Instantaneous Achievable Rates for the Three-User Case}
Similar to the previous subsection, we consider a specific channel realization, where the channel coefficients of the three users are given by  ${{\left| {{h_1}} \right|}^2}=0.5$, ${{\left| {{h_2}} \right|}^2}=0.4$, and ${{\left| {{h_3}} \right|}^2}=0.1$, respectively.
We are interested in the achievable rate vs. the received SNR at the BS, which is defined by $\frac{{\sum\nolimits_{k = 1}^K {{{\left| {{h_k}} \right|}^2}{P_k}} }}{{{N_0}}} = \frac{P}{{{N_0}}}$.
%${\rm SNR} \buildrel \Delta \over = \frac{{\sum\nolimits_{k = 1}^K {{{\left| {{h_k}} \right|}^2}{P_k}} }}{{{N_0}}} = \frac{P}{{{N_0}}}$.

%Fig.~\ref{aNOMA050401} shows the achievable rates of each user for the aNOMA scheme with $\beta=0.3$. As can be observed from the figure, the achievable rates of the aNOMA scheme are bounded by the derived upper- and lower-bounds, which is consistent with our analysis.
%Furthermore, the achievable rate of the conventional synchronized NOMA scheme is almost the same as the derived lower-bound. Therefore, it can be shown that the aNOMA scheme enjoys a greater achievable rate region compared to the conventional synchronized NOMA scheme. However, the improvement of the achievable rate due to the asynchronous transmission is not significant.
%
%\begin{figure}
%\centering
%\includegraphics[width=0.7\textwidth]{Fig/aNOMA050401.eps}
%\caption{The instantaneous achievable rate of aNOMA scheme, where three users with equal power allocation are considered. The channel coefficients are given by ${{\left| {{h_1}} \right|}^2}=0.5$, ${{\left| {{h_2}} \right|}^2}=0.4$, and ${{\left| {{h_3}} \right|}^2}=0.1$, respectively. The dash-dotted lines and the dashed lines are the derived upper- and lower-bounds, respectively.
%The transmitter shaping pulse is the RRC pulse with $\beta=0.3$.}
%\label{aNOMA050401}
%\centering
%\vspace{-6mm}
%\end{figure}

\begin{figure}
\centering
\includegraphics[width=0.6\textwidth]{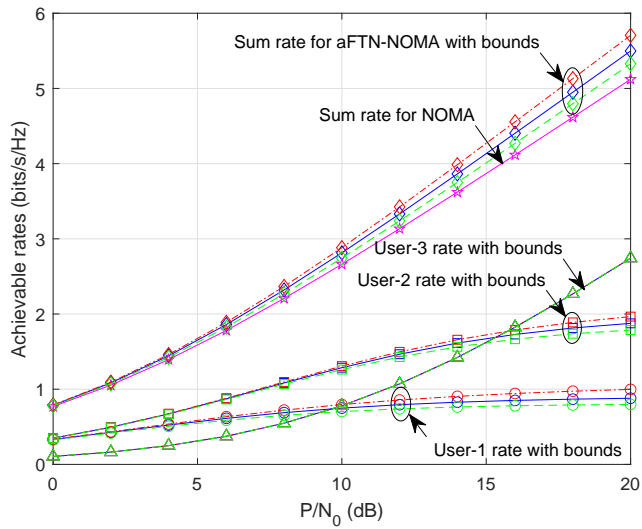}
\caption{The instantaneous achievable rate of aFTN-NOMA scheme for $\zeta=0.95$, where three users with equal power allocation are considered. The channel coefficients are given by ${{\left| {{h_1}} \right|}^2}=0.5$, ${{\left| {{h_2}} \right|}^2}=0.4$, and ${{\left| {{h_3}} \right|}^2}=0.1$, respectively. The red dash-dotted lines and the green dashed lines are the upper- and lower-bounds derived, while the blue solid line is the actual achievable rate.
The signaling pulse is the RRC pulse with $\beta=0.3$.}
\label{aFTN_NOMA050401_tau095}
\centering
\vspace{-8mm}
\end{figure}
Fig.~\ref{aFTN_NOMA050401_tau095} shows the achievable rates of each user for the aFTN-NOMA scheme with $\zeta=0.95$, and $\beta=0.3$. As shown in the figure, the achievable rates of the aFTN-NOMA scheme are perfectly bounded by the upper- and lower-bounds derived. On the other hand, we observe that asynchronous FTN transmission indeed attains rate improvements. Specifically, there is a $1.3$ dB gain for the aFTN-NOMA scheme with $\zeta=0.95$ compared to the conventional synchronous NOMA scheme at a rate of $5$ bits/s/Hz.
%Meanwhile, compared to the previous figure, we can see that the derived bounds become tighter with a smaller $\zeta$, this is consistent with our claim in Proposition~1. More details on this observation will be demonstrated in the following figure.

\begin{figure}
\centering
\includegraphics[width=0.6\textwidth]{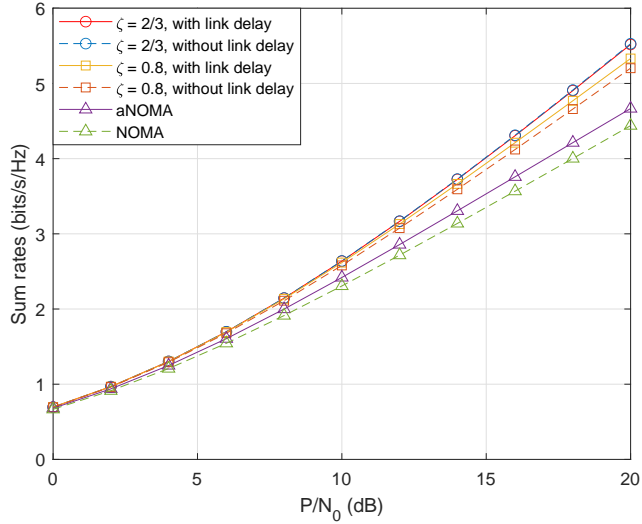}
\caption{An illustration of the trade-off between the SINR gain and the DoF gain for aFTN-NOMA schemes with different values of $\zeta$, where three users with equal power allocation are considered. The channel coefficients are given by ${{\left| {{h_1}} \right|}^2}=0.5$, ${{\left| {{h_2}} \right|}^2}=0.4$, and ${{\left| {{h_3}} \right|}^2}=0.1$, respectively. The signaling pulse is the RRC pulse with $\beta=0.5$.}
\label{Trade_off_simulation}
\centering
\vspace{-6mm}
\end{figure}
Let us demonstrate the SINR vs. DoF trade-off with $\beta=0.5$ in Fig.~\ref{Trade_off_simulation}, where we show the sum-rates of the aFTN-NOMA scheme, the aNOMA scheme, and the conventional synchronous NOMA scheme. For comparison, we also provide the corresponding synchronous rates without link delays. Specifically, we consider three cases at different symbol rates, i.e., $\zeta=1$, $\zeta=0.8$, and $\zeta=2/3$, respectively. As indicated in Proposition~1, the SINR gain reduces with the reduction of $\zeta$ from $\zeta=1$ to $\zeta=\frac{1}{{1 + \beta }} = \frac{2}{3}$, while the DoF gain increases. As observed from the figure, the SINR gain indeed decreases (corresponding to the shift along the x-axis) for a smaller $\zeta$, while the DoF gain (corresponding to the slope of the curves) increases. In particular, we notice that with $\zeta=2/3$, the curves of the aFTN-NOMA schemes operating with or without link delay are almost overlapped, which indicates that the SINR gain due to the link delay vanishes in this case. The observations confirm the accuracy of our derivations in Section III.

\subsection{Normalized Ergodic Achievable Rates for Single-Cell Setups}
In this subsection, we focus our attention on the ergodic achievable rates of both aFTN-NOMA and aNOMA, and compare them to that of the conventional power-domain NOMA in a single-cell system. In particular, we adopt the system settings from~\cite{WeiPerformanceGain} for our simulations, where we assume that the cell is modeled by a pair of concentric ring-shaped discs. The inner radius $D_0$ is introduced to model the minimum propagation path loss~\cite{WeiPerformanceGain}, while the outer radius $D_1$ represents the cell size. Furthermore, we assume that the BS is located at the center of the cell and all the users are uniformly scattered within the two concentric ring-shaped discs. To characterize the system's SNR, we adopt the definitions from~\cite{WeiPerformanceGain}, where the total average received SNR of all the users at the BS is defined by
\begin{align}
{\rm{SN}}{{\rm{R}}_{{\rm{sum}}}} \buildrel \Delta \over = \frac{{{P_{\max }}}}{{{N_0}}}\overline {{{\left| h \right|}^2}} .\label{ergodic_SNR_total}
\end{align}
In~\eqref{ergodic_SNR_total}, ${\overline {\left| h \right|^2}}$ denotes the average channel power gain with respect to the cell size and path loss model and it is calculated based on Equation (12) of~\cite{WeiPerformanceGain}. The term $P_{\max}$ in~\eqref{ergodic_SNR_total} denotes the total transmit power of the BS, which is adjusted adaptively for different cell sizes to provide the required ${\rm{SN}}{{\rm{R}}_{{\rm{sum}}}}$. Meanwhile, the noise PSD in~\eqref{ergodic_SNR_total} is set to be $N_0=-80$ dBm.
For reference, we summarize the related parameters in Table II.

\begin{table}[htbp]
\caption{Related Parameters for Simulations }
\centering
\begin{tabular}{|l|r|}
\hline
Packet length $N$~&~$100$ \\
\hline
Nyquist symbol period $T$~&~$1$ \\
\hline
Maximum value of link delay~&~$2T$ \\
\hline
Inner cell radius $D_0$~&~$50$ $\rm{m}$ \\
\hline
Outer cell radius $D_1$~&~$[75, 100, 200, 300,400,500]$ $\rm{m}$ \\
\hline
Number of users~&~ $[2, 4, 8, 16, 32, 64, 128]$\\
%\hline
%User density~&~ $[2, 4, 8, 16, 32, 64, 128]$ Users per ${{\rm{m}}^2}$\\
\hline
Path loss exponent $\alpha$~&~ $3.76$\\
\hline
Noise PSD $N_0$~&~$-80$ dBm\\
\hline
%System SNR ${\rm{SN}}{{\rm{R}}_{{\rm{sum}}}}$~&~$[0, 10, 20]$ dB\\
%\hline
\end{tabular}
\label{single_cell_channel_parameters}
\end{table}

\begin{figure}[htbp]
\centering
\subfigure[Normalized sum-rate vs. number of users with different SNRs.]{
\begin{minipage}[t]{0.5\textwidth}
\centering
\includegraphics[scale=0.5]{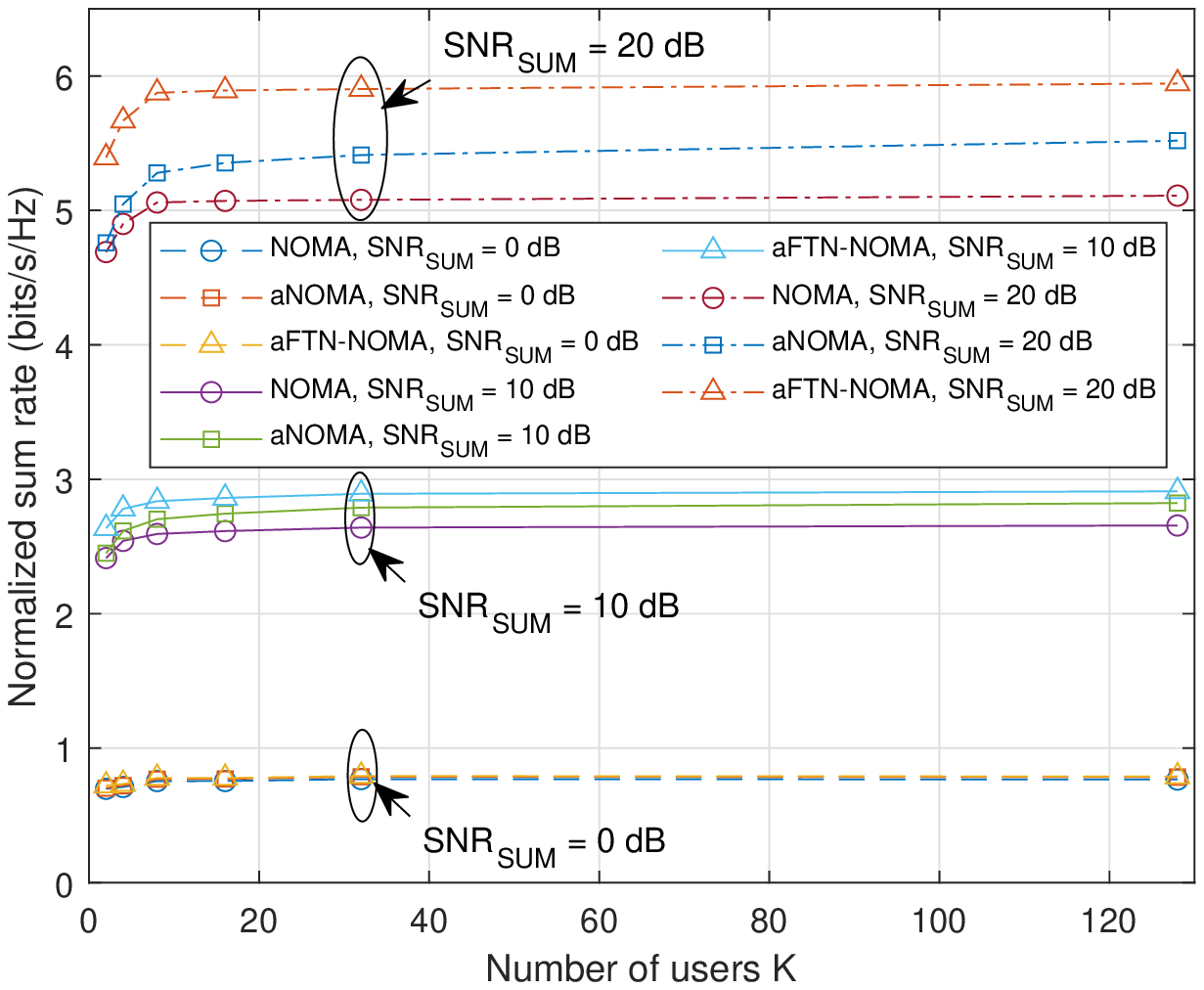}
%\caption{fig1}
\label{sum_rate_A}
\end{minipage}%
}%
\subfigure[Normalized sum-rate vs. cell range.]{
\begin{minipage}[t]{0.5\textwidth}
\centering
\includegraphics[scale=0.5]{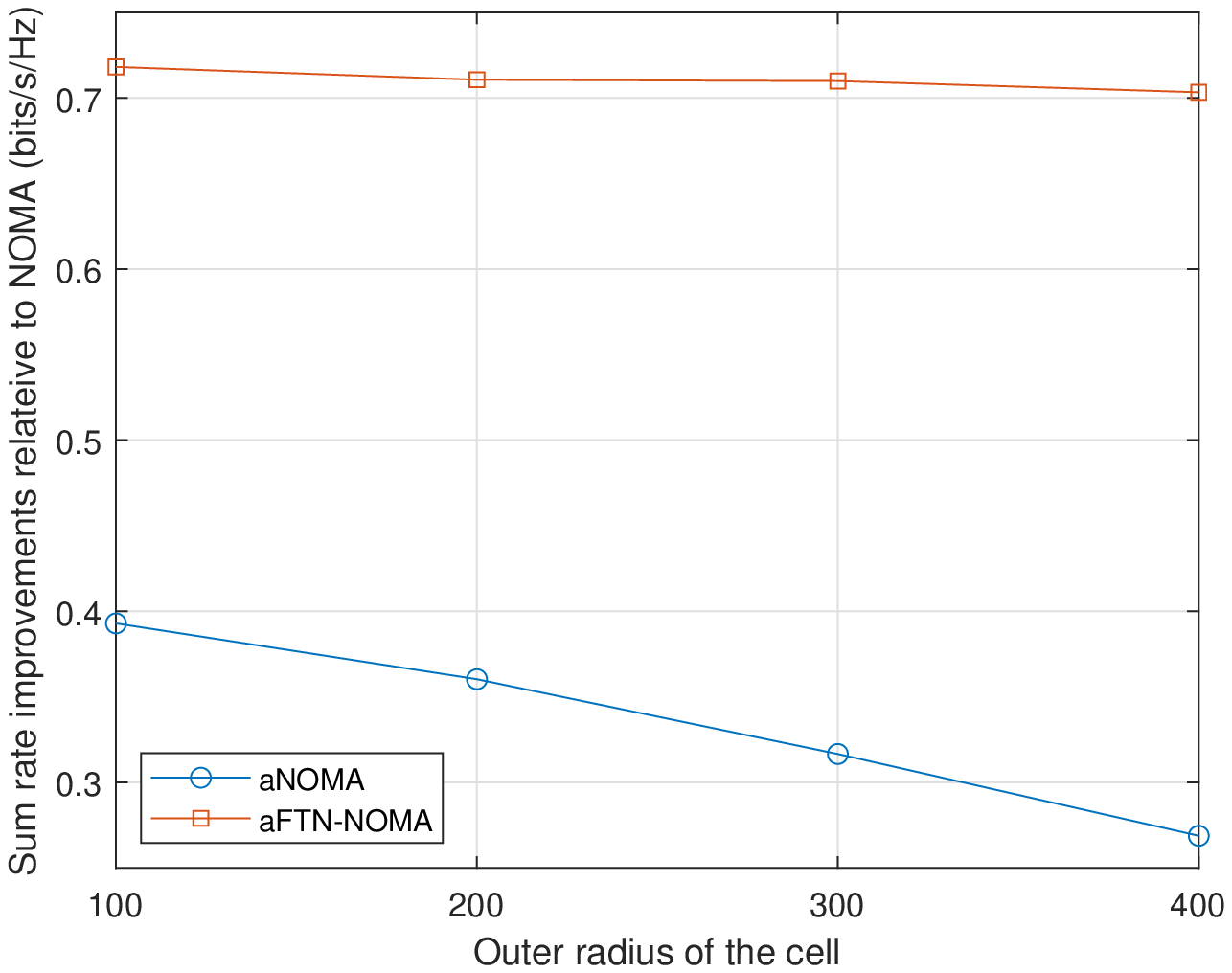}
%\caption{fig2}
\label{sum_rate_B}
\end{minipage}%
}%
\centering
\caption{Ergodic sum-rate analysis for the considered schemes, where the related simulation parameters are given in Table~\ref{single_cell_channel_parameters}.}
\label{System_simulation_A}
\end{figure}

We show the ergodic sum-rates of both aFTN-NOMA and aNOMA in comparison to that of NOMA in Fig.~\ref{sum_rate_A} versus ${\rm{SN}}{{\rm{R}}_{{\rm{sum}}}}$ and the numbers of users, where $\beta=0.3$, $\tau=0.75$, and $D_1=75$. As observed from Fig. 6(a), the ergodic sum-rates of both aFTN-NOMA and aNOMA are higher than the synchronous NOMA benchmark for various system SNRs, where the sum-rate improvements increase with ${\rm{SN}}{{\rm{R}}_{{\rm{sum}}}}$.
Indeed, both the DoF and the MUI are dominant factors in determining the achievable rates in the high-SNR regime. Therefore, FTN signaling and asynchronous transmission lead to beneficial rate improvements due to the exploitations of DoF and the mitigation of MUI.
On the other hand, we also observe from Fig.~\ref{sum_rate_A} that aNOMA achieves a higher sum-rate improvement for more users, while the sum-rate gap between aFTN-NOMA and NOMA is relatively constant for more than $8$ users. This is because the rate improvement of aNOMA arises from the MUI mitigation, while the rate improvement of aFTN-NOMA comes from its DoF gain. As the MUI increases with more users in the cell, the MUI mitigation leads to an increased rate improvement. However, as the DoF gain of aFTN-NOMA comes from the excess bandwidth of the signaling pulse, which does not change with the number of users in the cell, the rate improvement remains relatively constant for different number of users.

We portray the performance comparisons among NOMA, aNOMA, and aFTN-NOMA in terms of the cell size in Fig.~\ref{sum_rate_B}, where $\beta=0.3$, $\tau=0.75$, $K=128$, and ${\rm{SN}}{{\rm{R}}_{{\rm{sum}}}}=20$ dB. As indicated in Fig.~\ref{sum_rate_B}, the sum-rate gap between aFTN-NOMA and NOMA is relatively constant, while the sum-rate improvement of aNOMA is reduced for large cells. This observation is not unexpected because the DoF gain of aFTN-NOMA does not change with the channel characteristics. However, with the transmitted power fixed, the MUI
generally reduces for larger cells and therefore, the rate improvement of aNOMA also reduces.

\begin{figure}[htbp]
\centering
\subfigure[Sum rate vs. roll-off factors.]{
\begin{minipage}[t]{0.5\textwidth}
\centering
\includegraphics[scale=0.5]{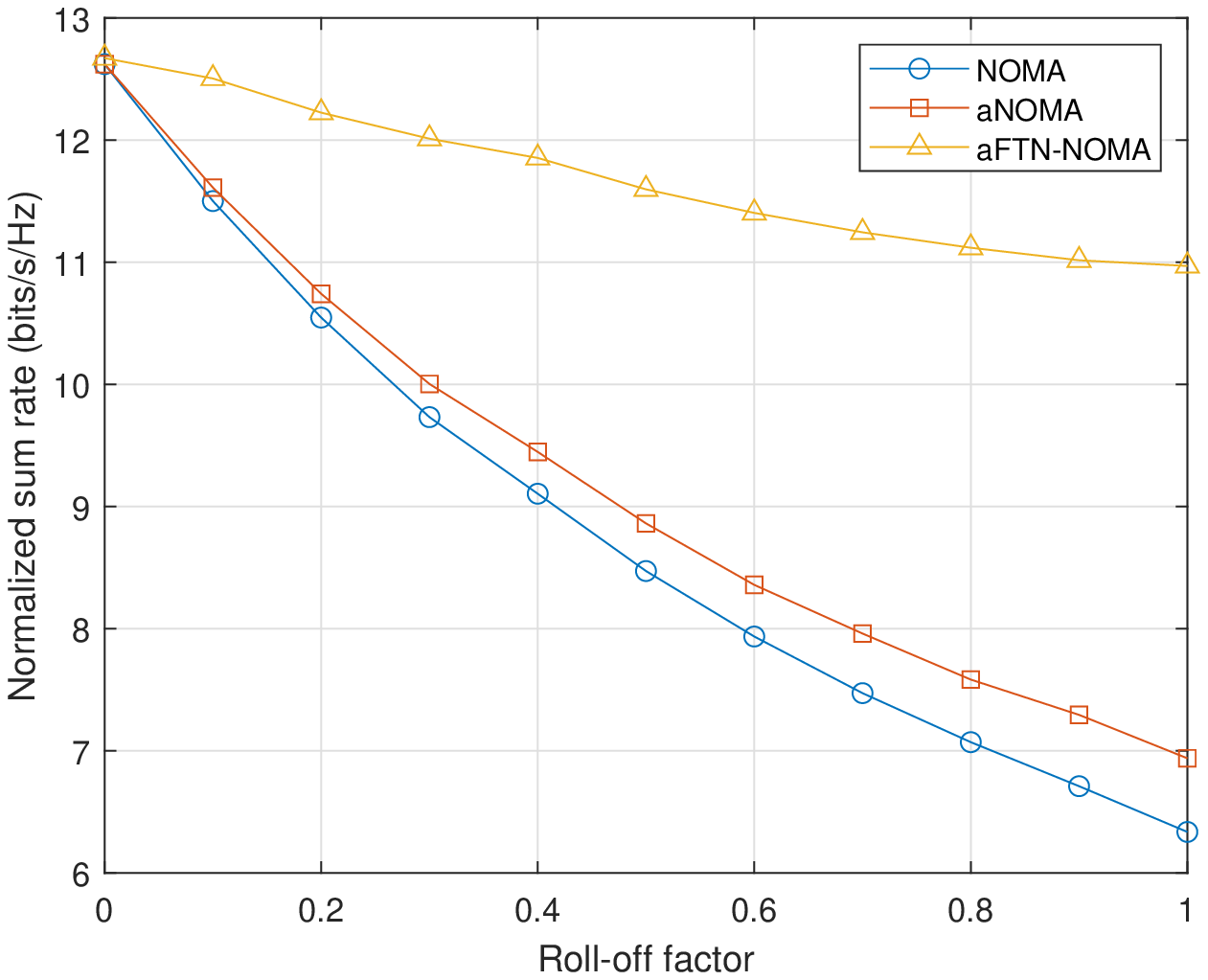}
%\caption{fig2}
\label{sum_rate_C}
\end{minipage}%
}%
\centering
\subfigure[CCDFs for different schemes.]{
\begin{minipage}[t]{0.5\textwidth}
\centering
\includegraphics[scale=0.5]{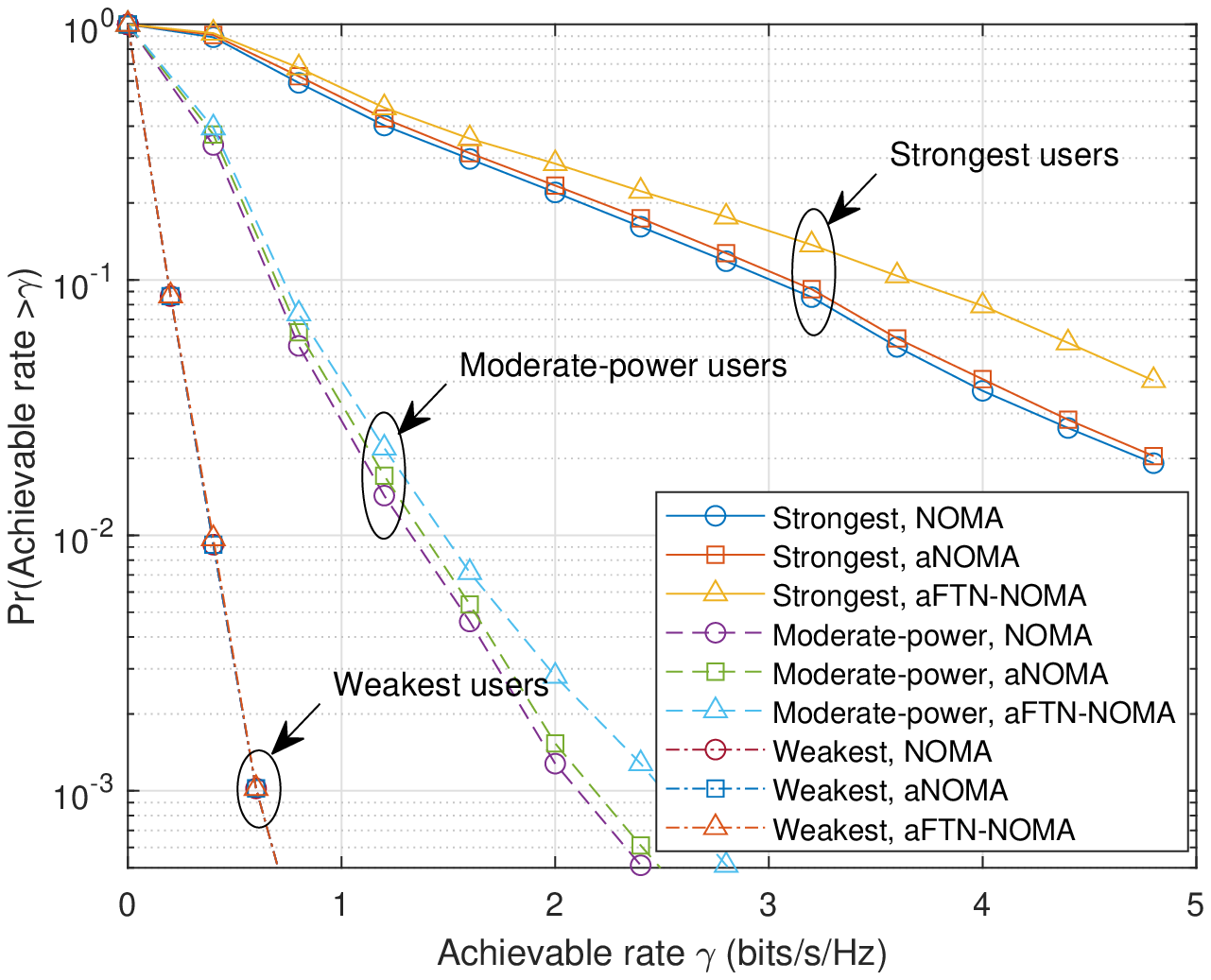}
%\caption{fig2}
\label{user_CDF}
\end{minipage}%
}%user_CDF.eps
\caption{Achievable rate analysis for the considered schemes, where the related simulation parameters are given in Table~\ref{single_cell_channel_parameters}.}
\label{System_simulation_B}
\end{figure}

The sum-rate performances versus roll-off factors are demonstrated in Fig.~\ref{sum_rate_C}, where $K=16$, $D_1=75$, and ${\rm{SN}}{{\rm{R}}_{{\rm{sum}}}}=40$ dB. In particular, we set $\tau=0.5$ for aFTN-NOMA system as it is sufficient to obtain the full DoF gains for $\beta  \in \left[ {0,1} \right]$. We observe that the sum-rate of all three schemes decreases for a larger $\beta$ due to the bandwidth normalization. However, the sum-rates of both NOMA and aNOMA reduce significantly compared to that of the aFTN-NOMA. This is because aFTN-NOMA efficiently exploits the DoF gain based on the excess bandwidth, which is consistent with our analysis in Section III-B. On the other hand, the sum-rate gap between NOMA and aNOMA increases for a larger roll-off factor. This is due to the fact that a larger excess bandwidth offers a stronger interference mitigation capability for aNOMA, which is also consistent with our discussions in Section III-A.

The comparison of complementary cumulative distribution functions (CCDFs) of different users is given in Fig.~\ref{user_CDF}, where $K=8$, $\beta=0.3$, $\tau=0.75$, $D_1=500$, and ${\rm{SN}}{{\rm{R}}_{{\rm{sum}}}}=20$ dB. Specifically, we show the CCDFs of the strongest user, a moderate-power user, and the weakest user in the schemes considered. As indicated from the figure, the CCDFs of both aNOMA and aFTN-NOMA show better performance than that of NOMA for both the strongest user and moderate-power user. However, the CCDFs of the weakest users are almost the same for the three schemes.
This is because the weakest user generally has low received SNR while the MUI is eliminated thanks to the SIC detection. In this case, neither the MUI mitigation, nor the DoF gain may be able to offer large rate improvements. Consequently, the performances of aFTN-NOMA, aNOMA, and NOMA are similar.

\section{Conclusions}
In this paper, we investigated the aNOMA and aFTN-NOMA schemes with the objective of improving the achievable rate of conventional NOMA transmissions. Specifically, we derived the corresponding achievable rate upper- and lower-bounds by invoking Szeg\"o's Theorem. We showed that asynchronous transmissions result in an SINR gain, while increasing the symbol rate may potentially lead to a DoF gain. More importantly, we unveiled that the associated SINR vs. DoF gain trade-off. In particular, we also showed that the SINR or DoF gains are related to the potential spectral aliasing and discussed the connections between the effect of spectral aliasing and the link delay as well as the symbol rate.
Our simulation results agreed with our analysis and demonstrated a significant achievable rate gain compared to the conventional NOMA transmission.

\appendices
\section{Proof of Lemma 1}
According to the chain rule and the calculation of the entropy of a multivariate normal distribution~\cite{cover2012elements}, we have
\begin{align}
{{I_{{\bf{h}},{\bm{\tau}},\zeta}}}\left( {{{\bf{y}}_k};{{\bf{x}}_k}| {{\bf{x}}_1}, \ldots ,{{\bf{x}}_{k-1}}} \right)
=& {h_{{\bf{h}},{\bm{\tau}},\zeta}\left( {{{\bf{y}}_k}\left| {{{\bf{x}}_1}, \ldots ,{{\bf{x}}_{k - 1}}} \right.} \right)}-{h_{{\bf{h}},{\bm{\tau}},\zeta}\left( {{{\bf{y}}_k}\left| {{{\bf{x}}_1}, \ldots ,{{\bf{x}}_{k}}} \right.} \right)}\notag\\
=&\frac{1}{2}\log_2 {\left( {2\pi e} \right)^N}\det \left( {\sum\limits_{l = k}^K {{{\left| {{h_l}} \right|}^2}{E_s}\left[ l \right]{{{\bf{\tilde G}}}_{l,k}}{\bf{\tilde G}}_{l,k}^{\rm{T}} + {N_0}{{{\bf{\tilde G}}}_{k,k}}} } \right)\notag\\
&\quad \!\!\!\!\!\!\!\! -\frac{1}{2}\log_2 {\left( {2\pi e} \right)^N}\det \left( {\sum\limits_{l = k+1}^K {{{\left| {{h_l}} \right|}^2}{E_s}\left[ l \right]{{{\bf{\tilde G}}}_{l,k}}{\bf{\tilde G}}_{l,k}^{\rm{T}} + {N_0}{{{\bf{\tilde G}}}_{k,k}}} } \right) .
\end{align}
Furthermore, according to the property of the matrix determinant, we have
\begin{align}
&{{I_{{\bf{h}},{\bm{\tau}},\zeta}}}\left( {{{\bf{y}}_k};{{\bf{x}}_k}| {{\bf{x}}_1}, \ldots ,{{\bf{x}}_{k-1}}} \right)\notag\\
=&\frac{1}{2}\log_2 \det \left[ {\left( {\sum\limits_{l = k}^K {{{\left| {{h_l}} \right|}^2}{E_s}\left[ l \right]{{{\bf{\tilde G}}}_{l,k}}{\bf{\tilde G}}_{l,k}^{\rm{T}} + {N_0}{{{\bf{\tilde G}}}_{k,k}}} } \right){{\left( {\sum\limits_{l = k + 1}^K {{{\left| {{h_l}} \right|}^2}{E_s}\left[ l \right]{{{\bf{\tilde G}}}_{l,k}}{\bf{\tilde G}}_{l,k}^{\rm{T}} + {N_0}{{{\bf{\tilde G}}}_{k,k}}} } \right)}^{ - 1}}} \right]\notag\\
=&\frac{1}{2}\log_2 \det \left[ {{{\bf{I}}_{N \times N}} + \frac{{{{\left| {{h_k}} \right|}^2}{E_s}\left[ k \right]{{{\bf{\tilde G}}}_{k,k}}{\bf{\tilde G}}_{k,k}^{\rm{T}}}}{{{N_0}}}{{\left( {{{{\bf{\tilde G}}}_{k,k}} + \frac{{\sum\limits_{l = k + 1}^K {{{\left| {{h_l}} \right|}^2}{E_s}\left[ l \right]} }}{{{N_0}}}{{{\bf{\tilde G}}}_{l,k}}{\bf{\tilde G}}_{l,k}^{\rm{T}}} \right)}^{ - 1}}} \right] .       \label{Lemma_det_der3}
\end{align}
This completes the proof of Lemma 1.
\section{Proof of Lemma 3}
According to the definition of ${{\bf{T}}_{l,k}}$, to verify the positive definiteness of ${{\bf{T}}_{l,k}}$ is equivalent to verify that ${\tilde{\bf{G}}_{l,k}}$ has a positive determinant.
It can be observed from~\eqref{G_lk_aFTN_NOMA} that ${\tilde{\bf{G}}_{l,k}}$ is a Gram matrix of nonzero energy functions $p\left( {t + n\zeta T + \tau \left[ l \right] - \tau \left[ k \right]} \right)$, for $n=0,\ldots, N-1$,
where the element of the $i$-th row and $j$-th column is given by the inner product of $p\left( {t + i\zeta T + \tau \left[ l \right] - \tau \left[ k \right]} \right)$ and $p\left( {t + j \zeta T + \tau \left[ l \right] - \tau \left[ k \right]} \right)$.
Note that the determinant of ${\tilde{\bf{G}}_{l,k}}$ is non-negative~\cite{gantmakher1959theory}. Therefore, we only have to verify that ${\tilde{\bf{G}}_{l,k}}$ has a nonzero determinant.
Gram's criterion \cite{gantmakher1959theory} indicates that for a set of strictly bandlimited functions $p\left( {t + n \zeta T + \tau \left[ l \right] - \tau \left[ k \right]} \right)$ having a finite energy in their frequency interval, the corresponding Gram matrix ${\tilde{\bf{G}}_{l,k}}$ has a nonzero determinant if and only if the set of functions $p\left( {t + n \zeta T + \tau \left[ l \right] - \tau \left[ k \right]} \right)$, for $n=0,\ldots, N-1$ are linearly independent.
In order to prove the linear independence, we consider Proposition 5.1.1 of \cite{brockwell1991time}, which indicates that a sufficient condition for the function set to be linearly independent is that
\begin{align}
\mathop {\lim }\limits_{n \to \infty } {\tilde g}\left[ {n,\tau \left[ l \right] - \tau \left[ k \right]} \right] = 0, \label{limit_g}
\end{align}
for any $1 \le l,k \le K$ \cite{kim2016properties}. Upon recalling~\eqref{IUI_aFTN_NOMA}, to prove~\eqref{limit_g}, we have to verify
\begin{align}
\mathop {\lim }\limits_{n \to \infty } \int\limits_{ - \infty }^\infty  {{{\left| {{H_p}\left( f \right)} \right|}^2}\exp \left( { j2\pi f\left( {n \zeta T + \tau \left[ l \right] - \tau \left[ k \right]} \right)} \right)} {\rm{d}}f = 0. \label{limit_g_expand}
\end{align}
Note that ${\tau \left[ l \right] - \tau \left[ k \right]}$ is constant for any given $l$ and $k$. Therefore, \eqref{limit_g_expand} holds due to the Riemann-Lebesgue Lemma \cite{rudin2006real}.
Furthermore, since ${\tilde{\bf{G}}_{l,k}}$ is a Toeplitz matrix, ${{\bf{T}}_{l,k}}$ is asymptotically a Toeplitz matrix, due to the fact that the product of two Toeplitz matrices is also asymptotically Toeplitz~\cite{gray2006toeplitz}.
On the other hand, it can be shown that the summation of two Toeplitz matrices is also a Toeplitz matrix, while the inverse of a Toeplitz matrix is asymptotically a Toeplitz matrix~\cite{gray2006toeplitz}. Therefore, it can be shown that ${\bf P}_k$ is asymptotically a Toeplitz matrix.
Meanwhile, we have verified that ${\tilde{\bf{G}}_{l,k}}$ is a positive definite Toeplitz matrix. Then, it can be shown that ${\bf P}_k$ is also positive definite, because the product of positive definite matrices is also positive definite.
This completes the proof of Lemma~3.

\section{Proof of Lemma 4}
As we know that the spectrum of RRC pulse is ${\left| {{H_p}\left( f \right)} \right|^2}$ strictly non-negative, we have
\begin{align}
\sum\limits_{k =  - \infty }^\infty  {{{\left| {{H_p}\left( {f - \frac{k}{\zeta T}} \right)} \right|}^2}} {e^{ - j2\pi \gamma k}} \le\sum\limits_{k =  - \infty }^\infty  {{{\left| {{H_p}\left( {f - \frac{k}{\zeta T}} \right)} \right|}^2}} {\left|{e^{ - j2\pi \gamma k}}\right|}
={\left| {{H_{{\rm{fo}}}}\left( f \right)} \right|^2} ,\label{Lemma4_der1}
\end{align}
where the bound becomes exact if $\gamma=0$.
On the other hand, notice that ${\left| {{H_p}\left( f \right)} \right|^2}$ is strictly bandlimited within the frequency interval $\left| f \right| \le \frac{{1 + \beta }}{{2T}}$. Thus, for $f \in \left[ { - \frac{1}{{2\zeta T}},\frac{1}{{2 \zeta T}}} \right]$, we have
\begin{align}
\sum\limits_{k =  - \infty }^\infty  {{{\left| {{H_p}\left( {f - \frac{k}{\zeta T}} \right)} \right|}^2}} {e^{ - j2\pi \gamma k}}
=&{\left| {{H_p}\left( f \right)} \right|^2} + {\left| {{H_p}\left( {f - \frac{1}{\zeta T}} \right)} \right|^2}{e^{ - j2\pi \gamma }} + {\left| {{H_p}\left( {f + \frac{1}{\zeta T}} \right)} \right|^2}{e^{j2\pi \gamma }} \notag\\
\ge &{\left| {{H_p}\left( f \right)} \right|^2} - {\left| {{H_p}\left( {f - \frac{1}{\zeta T}} \right)} \right|^2} - {\left| {{H_p}\left( {f + \frac{1}{\zeta T}} \right)} \right|^2} ,\notag\\
=&{\left| {{H_{{\rm{tfo}}}}\left( f \right)} \right|^2}.\label{Lemma4_der2}
\end{align}

Next, we discuss the achievability of the derived bounds. For $\beta=0$, it is not hard to notice that both ${\left| {{H_{{\rm{tfo}}}}\left( f \right)} \right|^2}$ and ${\left| {{H_{{\rm{fo}}}}\left( f \right)} \right|^2}$ become the same as the \emph{sinc} spectrum ${\left| {H_{\rm sinc}\left( f \right)} \right|^2}$. In this case, both the upper-bound of~\eqref{Lemma4_der1} and the lower-bound of~\eqref{Lemma4_der2} are achieved.
Furthermore, we note that both ${\left| {{H_{{\rm{fo}}}}\left( f \right)} \right|^2}$ and ${\left| {{{ H}_{{\rm{tfo}}}}\left( f \right)} \right|^2}$ become the exact RRC spectrum ${\left| {H_p\left( f \right)} \right|^2}$, when $\zeta \le 1/(1+\beta)$. Thus, both the upper-bound of~\eqref{Lemma4_der1} and the lower-bound of~\eqref{Lemma4_der2} are also achieved in this case.
This completes the proof of Lemma~4.

\section{Proof of Theorem 1}
Upon recalling~\eqref{IUI_aFTN_NOMA}, we have
%\begin{align}
%{{\tilde G}_{l,k}}\left( {2\pi f\zeta T} \right) &= \sum\limits_{n =  - \infty }^\infty  {{{\tilde g}_\zeta }\left[ {n,\Delta \tau } \right]{e^{ - j2\pi n\zeta Tf}}} \notag\\
%&= \sum\limits_{n =  - \infty }^\infty  {\int_{ - \infty }^\infty  {{{\left| {{H_p}\left( \lambda  \right)} \right|}^2}{e^{j2\pi \lambda \left( {n\zeta T + \Delta \tau } \right)}}} {\rm{d}}\lambda {e^{ - j2\pi n\zeta Tf}}} \notag\\
%&= \int_{ - \infty }^\infty  {{{\left| {{H_p}\left( \lambda  \right)} \right|}^2}} {e^{j2\pi \lambda \Delta \tau }}\sum\limits_{n =  - \infty }^\infty  {{e^{j2\pi n\zeta T\left( {\lambda  - f} \right)}}{\rm{d}}\lambda } .\label{The1_der1}
%\end{align}
\begin{align}
{{\tilde G}_{l,k}}\left( {2\pi f\zeta T} \right) = \sum\limits_{n =  - \infty }^\infty  {{{\tilde g}_\zeta }\left[ {n,\Delta \tau } \right]{e^{ - j2\pi n\zeta Tf}}}
= \int_{ - \infty }^\infty  {{{\left| {{H_p}\left( \lambda  \right)} \right|}^2}} {e^{j2\pi \lambda \Delta \tau }}\sum\limits_{n =  - \infty }^\infty  {{e^{j2\pi n\zeta T\left( {\lambda  - f} \right)}}{\rm{d}}\lambda } .\label{The1_der1}
\end{align}
Substituting the Poisson summation formula, i.e., $\sum\nolimits_{k =  - \infty }^\infty  {{e^{j2\pi kx}}}  = \sum\nolimits_{k=  - \infty }^\infty  {\delta \left( {x + k} \right)} $, into~\eqref{The1_der1} yields
\begin{align}
{{\tilde G}_{l,k}}\left( {2\pi f\zeta T} \right)
&= \int_{ - \infty }^\infty  {{{\left| {{H_p}\left( \lambda  \right)} \right|}^2}} {e^{j2\pi \lambda \Delta \tau }}\sum\limits_{n =  - \infty }^\infty  {\delta \left( {\zeta T\left( {\lambda  - f} \right) + n} \right){\rm{d}}\lambda } \notag\\
&=\frac{1}{{\zeta T}}\int_{ - \infty }^\infty  {{{\left| {{H_p}\left( \lambda  \right)} \right|}^2}} {e^{j2\pi \lambda \Delta \tau }}\sum\limits_{n =  - \infty }^\infty  {\delta \left( {\lambda  - \left( {f - \frac{n}{{\zeta T}}} \right)} \right){\rm{d}}\lambda } \label{The1_der2}\\
&= \frac{1}{{\zeta T}}\sum\limits_{n =  - \infty }^\infty  {{{\left| {{H_p}\left( {f - \frac{n}{{\zeta T}}} \right)} \right|}^2}{e^{j2\pi \left( {f - \frac{n}{{\zeta T}}} \right)\Delta \tau }}} ,\label{The1_der3}
\end{align}
where~\eqref{The1_der2} is due to the property of the Dirac delta function.
Therefore, by considering Lemma~4,~\eqref{The1_der3} can be bounded by
\begin{align}
\frac{1}{{\zeta T}}{\left| {{H_{{\rm{tfo}}}}\left( f \right)} \right|^2}\le {{\tilde G}_{l,k}}\left( {2\pi f\zeta T} \right) \le\frac{1}{{\zeta T}} {\left| {{H_{{\rm{fo}}}}\left( f \right)} \right|^2}.
\end{align}

It should be noted that the bounds derived are based on the conclusions from Lemma~4. Therefore, those bounds become exact when the conditions stated in Lemma~4 are satisfied.
This completes the proof of Theorem 1.

\section{Proof of Theorem 2}
Similar to the proof of Theorem 1, we substitute~\eqref{IUI_aFTN_NOMA} into~\eqref{DTFT_T_l_k} and obtain
\begin{align}
{{\tilde T}_{l,k}}\left( {2\pi f\zeta T} \right)
&= \sum\limits_{n =  - \infty }^\infty  {\sum\limits_{m =  - \infty }^\infty  {{{\tilde g}_\zeta }\left[ {m,\Delta \tau } \right]{{\tilde g}_\zeta }\left[ {m - n,\Delta \tau } \right]} } {e^{ - j2\pi n\zeta Tf}}\notag\\
&= \sum\limits_{m =  - \infty }^\infty  {{{\tilde g}_\zeta }\left[ {m,\Delta \tau } \right]} \int_{ - \infty }^\infty  {{{\left| {{H_p}\left( \lambda  \right)} \right|}^2}} {e^{j2\pi \lambda \left( {m\zeta T + \Delta \tau } \right)}}\sum\limits_{n =  - \infty }^\infty  {{e^{ - j2\pi n\zeta T\left( {\lambda  + f} \right)}}} {\rm{d}}\lambda .
\end{align}
Upon considering the Poisson summation formula, we have
\begin{align}
{{\tilde T}_{l,k}}\left( {2\pi f\zeta T} \right)
=& \sum\limits_{m =  - \infty }^\infty  {{{\tilde g}_\zeta }\left[ {m,\Delta \tau } \right]} \int_{ - \infty }^\infty  {{{\left| {{H_p}\left( \lambda  \right)} \right|}^2}} {e^{j2\pi \lambda \left( {m\zeta T + \Delta \tau } \right)}}\sum\limits_{n =  - \infty }^\infty  {\delta \left( {n - \zeta T\left( {\lambda  + f} \right)} \right)} {\rm{d}}\lambda\notag\\
=& \frac{1}{{\zeta T}}\sum\limits_{m =  - \infty }^\infty  {{{\tilde g}_\zeta }\left[ {m,\Delta \tau } \right]} {e^{j2\pi f\left( {m\zeta T + \Delta \tau } \right)}}\sum\limits_{n =  - \infty }^\infty  {{{\left| {{H_p}\left( {f - \frac{n}{{\zeta T}}} \right)} \right|}^2}} {e^{ - j2\pi mn - \frac{{n\Delta \tau }}{{\zeta T}}}}.\label{The2_der1}
\end{align}
Upon recalling Lemma~4,~\eqref{The2_der1} can be upper-bounded by
\begin{align}
{{\tilde T}_{l,k}}\left( {2\pi f\zeta T} \right) \le \frac{1}{{\zeta T}}{\left| {{H_{{\rm{fo}}}}\left( f \right)} \right|^2}\sum\limits_{m =  - \infty }^\infty  {{{\tilde g}_\zeta }\left[ {m,\Delta \tau } \right]} {e^{j2\pi f\left( {m\zeta T + \Delta \tau } \right)}}. \label{The2_der2}
\end{align}
Moreover, by considering Theorem 1,~\eqref{The2_der2} can be further upper-bounded by
\begin{align}
{{\tilde T}_{l,k}}\left( {2\pi f\zeta T} \right) \le {\left( {\frac{1}{{\zeta T}}{{\left| {{H_{{\rm{fo}}}}\left( f \right)} \right|}^2}} \right)^2}.
\end{align}
On the other hand,~\eqref{The2_der1} can be lower-bounded according to Lemma~4 by
\begin{align}
{{\tilde T}_{l,k}}\left( {2\pi f\zeta T} \right) \ge \frac{1}{{\zeta T}}{\left| {{H_{{\rm{tfo}}}}\left( f \right)} \right|^2}\sum\limits_{m =  - \infty }^\infty  {{{\tilde g}_\zeta }\left[ {m,\Delta \tau } \right]} {e^{j2\pi f\left( {m\zeta T + \Delta \tau } \right)}} .
\end{align}
Again, by considering Theorem 1,~\eqref{The2_der2} can be further lower-bounded by
\begin{align}
{{\tilde T}_{l,k}}\left( {2\pi f\zeta T} \right) \ge {\left( {\frac{1}{{\zeta T}}{{\left| {{H_{{\rm{tfo}}}}\left( f \right)} \right|}^2}} \right)^2}.
\end{align}

It should be noted that the bounds derived are based on the conclusions from Lemma~4. Therefore, those bounds become exact when the conditions stated in Lemma~4 are met.
This completes the proof of Theorem 2.
\section{Proof of Theorem 3}
We apply Szeg\"o's Theorem to~\eqref{Conditional_mutual_info_SIC} and obtain
\begin{align}
R_{{\bf{h}},{\bm{\tau}},\zeta}^{{k}} \buildrel \Delta \over = &\mathop {\lim }\limits_{N \to \infty } \frac{1}{N}{I_{{\bf{h}},\tau ,\zeta=1 }}\left( {{{\bf{y}}_k};{{\bf{x}}_k}|{{\bf{x}}_1}, \ldots ,{{\bf{x}}_{k - 1}}} \right) \notag\\
= &\frac{1}{{4\pi }}\int_{-\pi }^\pi  {{{\log }_2}\left( {1 + \frac{{{{\left| {{h_k}} \right|}^2}{E_s}\left[ k \right]{{\tilde T}_{k,k}}\left( \omega  \right)}}{{{N_0}}}{{\left( {{{\tilde G}_{k,k}}\left( \omega  \right) + \frac{{\sum\limits_{l = k + 1}^K {{{\left| {{h_l}} \right|}^2}{E_s}\left[ l \right]{{\tilde T}_{l,k}}\left( \omega  \right)} }}{{{N_0}}}} \right)}^{ - 1}}} \right)} {\rm{d}}\omega \notag\\
=&\frac{{\zeta T}}{2}\int_{ - \frac{1}{{2\zeta T}}}^{\frac{1}{{2\zeta T}}} {{{\log }_2}\left( {1 + \frac{{{{\left| {{h_k}} \right|}^2}{E_s}\left[ k \right]{{\tilde T}_{k,k}}\left( {2\pi \zeta Tf} \right)}}{{{N_0}{{\tilde G}_{k,k}}\left( {2\pi \zeta Tf} \right) + \sum\limits_{l = k + 1}^K {{{\left| {{h_l}} \right|}^2}{E_s}\left[ l \right]{{\tilde T}_{l,k}}\left( {2\pi \zeta Tf} \right)} }}} \right)} {\rm{d}}f. \label{The_3_der1}
\end{align}
Note that the integral range is within the frequency interval $f \in \left[ { - \frac{1}{{2 \zeta T}},\frac{1}{{2 \zeta T}}} \right]$. Therefore, the bounds in Theorem~1 and Theorem~2 can be applied for analysis. By observing that ${{\tilde G}_{k,k}}\left( {{2\pi \zeta Tf}} \right) = \frac{1}{\zeta T}{\left| {{H_{{\rm{fo}}}}\left( f \right)} \right|^2}$ and ${{\tilde T}_{k,k}}\left( {{2\pi \zeta Tf}} \right) = {\left( {\frac{1}{\zeta T}{{\left| {{H_{{\rm{fo}}}}\left( f \right)} \right|}^2}} \right)^2}$, we have
\begin{align}
R_{{\bf{h}},{\bm{\tau}},\zeta}^{{k}}
\le &\frac{{\zeta T}}{2}\int_{ - \frac{1}{{2\zeta T}}}^{\frac{1}{{2\zeta T}}} {{{\log }_2}\left( {1 + \frac{{{{\left| {{h_k}} \right|}^2}{E_s}\left[ k \right]{{\left( {\frac{{{{\left| {{H_{{\rm{fo}}}}\left( f \right)} \right|}^2}}}{{\zeta T}}} \right)}^2}}}{{{N_0}\frac{{{{\left| {{H_{{\rm{fo}}}}\left( f \right)} \right|}^2}}}{{\zeta T}} + \sum\limits_{l = k + 1}^K  {{\left| {{h_l}} \right|}^2}{E_s}\left[ l \right]{{\left( {\frac{{{{\left| {{H_{{\rm{tfo}}}}\left( f \right)} \right|}^2}}}{{\zeta T}}} \right)}^2}}}} \right)} {\rm{d}}f ,
\end{align}
which can be further simplified as
\begin{align}
R_{{\bf{h}},{\bm{\tau}},\zeta}^{{k}}
\le &\frac{{\zeta T}}{2}\int_{ - \frac{1}{{2\zeta T}}}^{\frac{1}{{2\zeta T}}} {{{\log }_2}\left( {1 + \frac{{{{\left| {{h_k}} \right|}^2}{P_k}{{\left| {{H_{{\rm{fo}}}}\left( f \right)} \right|}^2}}}{{{N_0} + \sum\limits_{l = k + 1}^K  {{\left| {{h_l}} \right|}^2}{P_l}{{\left| {{H_{{\rm{tfo}}}}\left( f \right)} \right|}^2}\rho\left( f \right)}}} \right)} {\rm{d}}f. \label{The_3_der2}
\end{align}
On the other hand, we have
\begin{align}
R_{{\bf{h}},{\bm{\tau}},\zeta}^{{k}}
\ge& \frac{{\zeta T}}{2}\int_{ - \frac{1}{{2\zeta T}}}^{\frac{1}{{2\zeta T}}} {{{\log }_2}\left( {1 + \frac{{{{\left| {{h_k}} \right|}^2}{E_s}\left[ k \right]{{\left( {\frac{{{{\left| {{H_{{\rm{fo}}}}\left( f \right)} \right|}^2}}}{{\zeta T}}} \right)}^2}}}{{{N_0}\frac{{{{\left| {{H_{{\rm{fo}}}}\left( f \right)} \right|}^2}}}{{\zeta T}} + \sum\limits_{l = k + 1}^K  {{\left| {{h_l}} \right|}^2}{E_s}\left[ l \right]{{\left( {\frac{{{{\left| {{H_{{\rm{fo}}}}\left( f \right)} \right|}^2}}}{{\zeta T}}} \right)}^2}}}} \right)} {\rm{d}}f \notag\\
=&\frac{{\zeta T}}{2}\int_{ - \frac{1}{{2\zeta T}}}^{\frac{1}{{2\zeta T}}} {{{\log }_2}\left( {1 + \frac{{{{\left| {{h_k}} \right|}^2}{P_k}{{\left| {{H_{{\rm{fo}}}}\left( f \right)} \right|}^2}}}{{{N_0} + \sum\limits_{l = k + 1}^K  {{\left| {{h_l}} \right|}^2}{P_l}{{\left| {{H_{{\rm{fo}}}}\left( f \right)} \right|}^2}}}} \right)} {\rm{d}}f. \label{The_3_der3}
\end{align}
Furthermore, by considering the symbol rate $\frac{1}{{\zeta T}}$, the signal dimension, and the signal bandwidth $2W$, the bounds in~\eqref{The_3_der2} and~\eqref{The_3_der3} can be normalized as shown in~\eqref{single_user_upper_bound_aFTN_NOMA} and~\eqref{single_user_lower_bound_aFTN_NOMA}.
Finally, it can be shown that the bounds derived become exact when the conditions stated in Lemma~4 are met.
This completes the proof of Theorem 3.

\bibliographystyle{IEEEtran}
\bibliography{Improved_NOMA}

\begin{thebibliography}{10}
\providecommand{\url}[1]{#1}
\csname url@samestyle\endcsname
\providecommand{\newblock}{\relax}
\providecommand{\bibinfo}[2]{#2}
\providecommand{\BIBentrySTDinterwordspacing}{\spaceskip=0pt\relax}
\providecommand{\BIBentryALTinterwordstretchfactor}{4}
\providecommand{\BIBentryALTinterwordspacing}{\spaceskip=\fontdimen2\font plus
\BIBentryALTinterwordstretchfactor\fontdimen3\font minus
  \fontdimen4\font\relax}
\providecommand{\BIBforeignlanguage}[2]{{%
\expandafter\ifx\csname l@#1\endcsname\relax
\typeout{** WARNING: IEEEtran.bst: No hyphenation pattern has been}%
\typeout{** loaded for the language `#1'. Using the pattern for}%
\typeout{** the default language instead.}%
\else
\language=\csname l@#1\endcsname
\fi
#2}}
\providecommand{\BIBdecl}{\relax}
\BIBdecl

\bibitem{ShuangyangWCNC}
S.~Li, Z.~Wei, W.~Yuan, J.~Yuan, B.~Bai, and D.~W.~K. Ng, ``On the achievable
  rates of uplink {NOMA} with asynchronized transmission,'' in \emph{Proc. IEEE
  Wireless Commun. Net. Conf.}, 2020, pp. 1--6.

\bibitem{ding2017survey}
Z.~Ding, X.~Lei, G.~K. Karagiannidis, R.~Schober, J.~Yuan, and V.~K. Bhargava,
  ``A survey on non-orthogonal multiple access for {5G} networks: Research
  challenges and future trends,'' \emph{IEEE J. Sel. Areas Commun.}, vol.~35,
  no.~10, pp. 2181--2195, Oct. 2017.

\bibitem{WeiPerformanceGain}
Z.~{Wei}, L.~{Yang}, D.~W.~K. {Ng}, J.~{Yuan}, and L.~{Hanzo}, ``On the
  performance gain of {NOMA} over {OMA} in uplink communication systems,''
  \emph{IEEE Trans. Commun.}, vol.~68, no.~1, pp. 536--568, Jan. 2020.

\bibitem{Wu2020Towards}
Q.~{Wu} and R.~{Zhang}, ``Towards smart and reconfigurable environment:
  Intelligent reflecting surface aided wireless network,'' \emph{IEEE Commun.
  Mag.}, vol.~58, no.~1, pp. 106--112, Nov. 2020.

\bibitem{wang2021minimum}
Y.~Wang, J.~Wang, D.~W.~K. Ng, R.~Schober, and X.~Gao, ``A minimum error
  probability {NOMA} design,'' \emph{IEEE Trans. Wireless Commun.}, Jul. 2021.

\bibitem{rimoldi1996rate}
B.~Rimoldi and R.~Urbanke, ``A rate-splitting approach to the {Gaussian}
  multiple-access channel,'' \emph{IEEE Trans. Inf. Theory}, vol.~42, no.~2,
  pp. 364--375, Mar. 1996.

\bibitem{Clerckx2016rate}
B.~{Clerckx}, H.~{Joudeh}, C.~{Hao}, M.~{Dai}, and B.~{Rassouli}, ``Rate
  splitting for {MIMO} wireless networks: A promising {PHY}-layer strategy for
  {LTE} evolution,'' \emph{IEEE Commun. Mag.}, vol.~54, no.~5, pp. 98--105, May
  2016.

\bibitem{Ding2020simple_IRS_NOMA}
Z.~Ding and H.~Vincent~Poor, ``A simple design of {IRS-NOMA} transmission,''
  \emph{IEEE Commun. Lett.}, vol.~24, no.~5, pp. 1119--1123, Feb. 2020.

\bibitem{Moltafet2018comparison}
M.~Moltafet, N.~M. Yamchi, M.~R. Javan, and P.~Azmi, ``Comparison study between
  {PD-NOMA} and {SCMA},'' \emph{IEEE Trans. Veh. Technol.}, vol.~67, no.~2, pp.
  1830--1834, Feb. 2018.

\bibitem{HaciAsyn}
H.~{Haci}, H.~{Zhu}, and J.~{Wang}, ``Performance of non-orthogonal multiple
  access with a novel asynchronous interference cancellation technique,''
  \emph{IEEE Trans. Commun.}, vol.~65, no.~3, pp. 1319--1335, Mar. 2017.

\bibitem{Zou2019analysis}
X.~{Zou}, B.~{He}, and H.~{Jafarkhani}, ``An analysis of two-user uplink
  asynchronous non-orthogonal multiple access systems,'' \emph{IEEE Trans.
  Wireless Commun.}, vol.~18, no.~2, pp. 1404--1418, Jan. 2019.

\bibitem{Yuan2020iterative}
W.~{Yuan}, N.~{Wu}, Q.~{Guo}, D.~W.~K. {Ng}, J.~{Yuan}, and L.~{Hanzo},
  ``Iterative joint channel estimation, user activity tracking, and data
  detection for {FTN-NOMA} systems supporting random access,'' \emph{IEEE
  Trans. Commun.}, vol.~68, no.~5, pp. 2963--2977, May 2020.

\bibitem{General-anderson2013faster}
J.~B. Anderson, F.~Rusek, and V.~{\"O}wall, ``Faster-than-{Nyquist}
  signaling,'' \emph{Proc. IEEE}, vol. 101, no.~8, pp. 1817--1830, Aug. 2013.

\bibitem{li2020code}
S.~{Li}, J.~{Yuan}, B.~{Bai}, and N.~{Benvenuto}, ``Code-based channel
  shortening for faster-than-{Nyquist} signaling: Reduced-complexity detection
  and code design,'' \emph{IEEE Trans. Commun.}, vol.~68, no.~7, pp.
  3996--4011, Jul. 2020.

\bibitem{sugiura2014frequency}
S.~Sugiura and L.~Hanzo, ``Frequency-domain-equalization-aided iterative
  detection of faster-than-{Nyquist} signaling,'' \emph{IEEE Trans. Veh.
  Technol.}, vol.~64, no.~5, pp. 2122--2128, May 2014.

\bibitem{ishihara2021evolution}
T.~Ishihara, S.~Sugiura, and L.~Hanzo, ``The evolution of faster-than-{Nyquist}
  signaling,'' \emph{\emph{to appear in} IEEE Access}, 2021.

\bibitem{kanaras2009spectrally}
I.~Kanaras, A.~Chorti, M.~R. Rodrigues, and I.~Darwazeh, ``Spectrally efficient
  {FDM} signals: Bandwidth gain at the expense of receiver complexity,'' in
  \emph{IEEE Int. Conf. Commun.}, 2009, pp. 1--6.

\bibitem{darwazeh2013optical}
I.~Darwazeh, T.~Xu, T.~Gui, Y.~Bao, and Z.~Li, ``Optical {SEFDM} system;
  bandwidth saving using non-orthogonal sub-carriers,'' \emph{IEEE Photonics
  Technol. Lett.}, vol.~26, no.~4, pp. 352--355, 2013.

\bibitem{li2018superposition}
S.~Li, B.~Bai, J.~Zhou, Q.~He, and Q.~Li, ``Superposition coded modulation
  based {faster-than-Nyquist} signaling,'' \emph{Wireless Commun. Mobile
  Comput.}, vol. 2018, 2018.

\bibitem{rusek2009constrained}
F.~Rusek and J.~B. Anderson, ``Constrained capacities for faster-than-{Nyquist}
  signaling,'' \emph{IEEE Trans. Inf. Theory}, vol.~55, no.~2, pp. 764--775,
  Feb. 2009.

\bibitem{Dai2018survey}
L.~{Dai}, B.~{Wang}, Z.~{Ding}, Z.~{Wang}, S.~{Chen}, and L.~{Hanzo}, ``A
  survey of non-orthogonal multiple access for {5G},'' \emph{IEEE Commun. Sur.
  \& Tut.}, vol.~20, no.~3, pp. 2294--2323, May 2018.

\bibitem{gray2006toeplitz}
R.~M. Gray, \emph{Toeplitz and Circulant Matrices: A Review}.\hskip 1em plus
  0.5em minus 0.4em\relax Now Foundations and Trends, 2006.

\bibitem{ding2014performance}
Z.~Ding, Z.~Yang, P.~Fan, and H.~V. Poor, ``On the performance of
  non-orthogonal multiple access in {5G} systems with randomly deployed
  users,'' \emph{IEEE Signal Process. Lett.}, vol.~21, no.~12, pp. 1501--1505,
  2014.

\bibitem{li2017reduced}
S.~Li, B.~Bai, J.~Zhou, P.~Chen, and Z.~Yu, ``Reduced-complexity equalization
  for {faster-than-Nyquist} signaling: New methods based on {Ungerboeck}
  observation model,'' \emph{IEEE Trans. Commun.}, vol.~66, no.~3, pp.
  1190--1204, Mar. 2017.

\bibitem{li2020time}
S.~Li, W.~Yuan, J.~Yuan, B.~Bai, D.~W.~K. Ng, and L.~Hanzo, ``Time-domain vs.
  frequency-domain equalization for {FTN} signaling,'' \emph{IEEE Trans. Veh.
  Technol.}, vol.~69, no.~8, pp. 9174--9179, Jun. 2020.

\bibitem{Zhiqiang2017optimal}
Z.~{Wei}, D.~W.~K. {Ng}, J.~{Yuan}, and H.~{Wang}, ``Optimal resource
  allocation for power-efficient {MC-NOMA} with imperfect channel state
  information,'' \emph{IEEE Trans. Commun.}, vol.~65, no.~9, pp. 3944--3961,
  May 2017.

\bibitem{Zhiqiang2019NOMAHybrid}
Z.~{Wei}, D.~W.~K. {Ng}, and J.~{Yuan}, ``{NOMA} for hybrid {mmWave}
  communication systems with beamwidth control,'' \emph{IEEE J. Sel. Topics
  Signal Process.}, vol.~13, no.~3, pp. 567--583, Jun. 2019.

\bibitem{Yuan2020Joint}
W.~{Yuan}, N.~{Wu}, J.~{Yuan}, D.~W.~K. {Ng}, and L.~{Hanzo}, ``Joint data and
  active user detection for grant-free {FTN-NOMA} in dynamic networks,'' in
  \emph{IEEE Int. Conf. Commun.}, 2020, pp. 1--6.

\bibitem{Yuan2020iterative_SCMA}
W.~{Yuan}, N.~{Wu}, A.~{Zhang}, X.~{Huang}, Y.~{Li}, and L.~{Hanzo},
  ``Iterative receiver design for {FTN} signaling aided sparse code multiple
  access,'' \emph{IEEE Trans. Wireless Commun.}, vol.~19, no.~2, pp. 915--928,
  Nov. 2020.

\bibitem{simon2010szegHo}
B.~Simon, \emph{{Szeg{\H{o}}'s Theorem and Its Descendants: Spectral Theory for
  L2 Perturbations of Orthogonal Polynomials}}.\hskip 1em plus 0.5em minus
  0.4em\relax Princeton university press, 2010, vol.~6.

\bibitem{qiu2019downlink}
M.~Qiu, Y.-C. Huang, and J.~Yuan, ``Downlink non-orthogonal multiple access
  without {SIC} for block fading channels: An algebraic rotation approach,''
  \emph{IEEE Trans. Wireless Commun.}, vol.~18, no.~8, pp. 3903--3918, Jun.
  2019.

\bibitem{qiu2018lattice}
M.~Qiu, Y.-C. Huang, J.~Yuan, and C.-L. Wang, ``Lattice-partition-based
  downlink non-orthogonal multiple access without {SIC} for slow fading
  channels,'' \emph{IEEE Trans. Commun.}, vol.~67, no.~2, pp. 1166--1181, Oct.
  2018.

\bibitem{kim2016properties}
Y.~J.~D. Kim, ``Properties of {faster-than-Nyquist} channel matrices and
  folded-spectrum, and their applications,'' in \emph{Proc. IEEE Wireless
  Commun. Net. Conf.}, 2016, pp. 1--7.

\bibitem{kim2016faster}
Y.~J.~D. Kim, J.~Bajcsy, and D.~Vargas, ``Faster-than-{Nyquist} broadcasting in
  {Gaussian} channels: Achievable rate regions and coding,'' \emph{IEEE Trans.
  Commun.}, vol.~64, no.~3, pp. 1016--1030, 2016.

\bibitem{rusek2011optimal}
F.~Rusek and A.~Prlja, ``Optimal channel shortening for {MIMO} and {ISI}
  channels,'' \emph{IEEE Trans. Wireless Commun.}, vol.~11, no.~2, pp.
  810--818, Dec. 2011.

\bibitem{Ameha2018iterative}
A.~T. Abebe and C.~G. Kang, ``Iterative decoders for {FTN-based} {NOMA} scheme
  to multiplex sporadic and broadband transmission,'' in \emph{Int. Conf. Inf.
  Commun. Techn. Conv. (ICTC)}, 2018, pp. 813--817.

\bibitem{cover2012elements}
T.~M. Cover and J.~A. Thomas, \emph{Elements of Information Theory}.\hskip 1em
  plus 0.5em minus 0.4em\relax John Wiley \& Sons, 2012.

\bibitem{Clerckx2021critical}
B.~Clerckx, Y.~Mao, R.~Schober, E.~A. Jorswieck, D.~J. Love, J.~Yuan, L.~Hanzo,
  G.~Y. Li, E.~G. Larsson, and G.~Caire, ``Is {NOMA} efficient in multi-antenna
  networks? {A} critical look at next generation multiple access techniques,''
  \emph{IEEE Open J. Commun. Soc.}, vol.~2, pp. 1310--1343, Jun. 2021.

\bibitem{gantmakher1959theory}
F.~R. Gantmakher, \emph{The Theory of Matrices}.\hskip 1em plus 0.5em minus
  0.4em\relax American Mathematical Soc., 1959, vol. 131.

\bibitem{brockwell1991time}
P.~J. Brockwell, R.~A. Davis, and S.~E. Fienberg, \emph{Time Series: Theory and
  Methods}.\hskip 1em plus 0.5em minus 0.4em\relax Springer Science \& Business
  Media, 1991.

\bibitem{rudin2006real}
W.~Rudin, \emph{Real and Complex Analysis}.\hskip 1em plus 0.5em minus
  0.4em\relax Tata McGraw-hill education, 2006.

\end{thebibliography}

\end{document}